\documentclass[conference,compsoc]{IEEEtran}

\usepackage{versions}
\usepackage{cite}
\usepackage{multirow}
\usepackage[inline]{enumitem}
\setlist{nolistsep}
\usepackage[colorlinks=true,citecolor=blue]{hyperref}
\usepackage{xurl}
\usepackage[dvipsnames]{xcolor}
\usepackage{graphicx}
\usepackage{tikz}
\usepackage{booktabs}
\usepackage{soul}
\usepackage{listings}
\usepackage{amsmath}
\usepackage{amssymb}
\usepackage{microtype}
\usetikzlibrary{shapes,arrows,arrows.meta,calc,decorations.pathmorphing}
\usetikzlibrary{backgrounds, positioning, fit}
\usetikzlibrary{positioning}
\tikzset{main node/.style={circle,fill=blue!20,draw,minimum size=0.75cm,inner sep=0pt}}
\usepackage{xspace}
\usepackage{array}
\newcolumntype{C}[1]{>{\centering\arraybackslash}p{#1}}

\usepackage{pifont}
\newcommand{\cmark}{\ding{51}\xspace}%
\newcommand{\xmark}{\ding{55}\xspace}%

\newcommand{\method}{\textsc{Knox}\xspace}

\newcommand{\titlename}{\method: Hardware-assisted Regulatory Sandbox for Generative Models}

\newcommand{\adv}{\protect{$\mathcal{A}dv$}\xspace}

\renewcommand\adv{\ensuremath{\sf{\mathcal Adv}}\xspace}

\definecolor{tdgreen}{RGB}{47,111,73}
\definecolor{tdgreenfill}{RGB}{235,245,238}
\definecolor{tdblue}{RGB}{47,84,120}
\definecolor{tdbluefill}{RGB}{235,241,247}
\definecolor{hwgray}{RGB}{115,115,110}
\definecolor{hwgrayfill}{RGB}{244,244,242}
\definecolor{enfgold}{RGB}{178,134,32}
\definecolor{enfgoldfill}{RGB}{252,243,219}
\definecolor{unred}{RGB}{164,72,66}
\definecolor{unredfill}{RGB}{250,238,236}
\definecolor{extpurple}{RGB}{95,80,134}
\definecolor{extpurplefill}{RGB}{240,237,246}

\newcommand{\traj}{\ensuremath{\mathcal{T}}}
\newcommand{\act}{\ensuremath{\alpha}}

\newcommand{\derive}{\ensuremath{\mathrm{derive}}}

\newcommand{\signatt}{\ensuremath{\mathsf{Sign}_{\mathrm{att}}}}

\newcommand{\modelverb}[2][blue!70]{%
  \textbf{\upshape\textcolor{#1}{\lstinline[breaklines=false, basicstyle=\ttfamily]{#2}}}%
}
\definecolor{jsonstr}{RGB}{23,110,96}
\definecolor{jsoncmt}{RGB}{130,130,130}
\definecolor{jsonkey}{RGB}{40,60,140}
\lstdefinelanguage{jsonc}{
  morestring=[b]",
  morecomment=[l]{//},
  sensitive=true,
  morekeywords={trajectory,position,agent,tool,params,hash,name,scope,predicate,
                spec,syscalls,paths,rlimits,regulations,derivation,budget,stop,
                max_turns,wall_s,command,to,path},
}
\lstdefinestyle{apm}{
  language=jsonc,
  mathescape=true,
  basicstyle=\ttfamily\footnotesize,
  stringstyle=\color{jsonstr},
  keywordstyle=\color{jsonkey}\bfseries,
  commentstyle=\color{jsoncmt}\itshape,
  showstringspaces=false,
  columns=fixed,
  basewidth=0.5em,
  keepspaces=true,
  frame=tb,
  framerule=0.7pt,
  rulecolor=\color{black!55},
  framesep=6pt,
  aboveskip=2pt,
  belowskip=2pt,
  xleftmargin=4pt,
  xrightmargin=4pt,
  breaklines=false,
}

\usepackage[ruled]{algorithm2e}
\usepackage{stfloats}
\usepackage{cleveref}             

\crefname{algocf}{attestation}{attestations}
\Crefname{algocf}{Attestation}{Attestations}
\crefname{algorithm}{attestation}{attestations}
\Crefname{algorithm}{Attestation}{Attestations}

\newenvironment{attdesc}[1][htb]
  {%
   \begin{algorithm}[#1]}{\end{algorithm}}

\newcommand{\RefMon}{AttRM\xspace}
\usepackage[flushmargin,hang]{footmisc}

\usepackage{amsthm}
\usepackage{thmtools}
\usepackage[most]{tcolorbox}

\usepackage{array}
\newcolumntype{C}[1]{>{\centering\arraybackslash}p{#1}}

\definecolor{blond}{rgb}{0.98, 0.94, 0.75} 

\declaretheoremstyle[numbered=no, headfont=\bfseries, postheadspace=0pt, headpunct={}]{thmsty} 

\declaretheorem[style=thmsty, name={}]{takeaway} 

\tcolorboxenvironment{takeaway}{enhanced jigsaw, colback=blond, drop shadow, boxrule=0.9pt, boxsep=0.1pt, left=4pt, right=4pt, top=4pt, bottom=4pt}

\includeversion{arxiv}
\excludeversion{submit}

\begin{document}

\title{\titlename}

\begin{arxiv}
\author{
    \IEEEauthorblockN{Vasisht Duddu$^{*\dagger}$, Adam Caulfield\textsuperscript{1}$^{\dagger}$, Xi He$^{*\dagger}$}
    \IEEEauthorblockA{$^{*}$Vector Institute, $^{\dagger}$University of Waterloo}
    \IEEEauthorblockA{vasisht.duddu@vectorinstitute.ai, acaulfield@uwaterloo.ca, xi.he@uwaterloo.ca}
} 
\end{arxiv}

\maketitle

\begin{abstract}
Regulations require companies to run AI models in a \emph{regulatory sandbox}, an isolated environment that enables monitoring model behavior by regulators, prior to deployment.
But, they do not spell out the technical mechanisms to design such a sandbox.
Ideally, a regulatory sandbox should show
\begin{enumerate*}[label={(\roman*)}]
\item \emph{compliance} (checks against regulations, and deviations from intended behavior, i.e., unauthorized tool and system calls);
\item \emph{confidentiality} (protects model weights from regulators);
\item \emph{attestability} (produces tamper-resistant evidence of model behavior for monitoring); and
\item \emph{scalability} (evidence is verifiable by multiple regulators).
\end{enumerate*}
Several prior works (information-flow control, prompt-injection defenses, control-flow integrity, and traditional sandboxing) address specific compliance aspects but lack confidentiality, attestability, and scalability. TEEs meet some of these requirements, but simply running models inside a TEE is insufficient as we need compliance mechanisms.

We present \method, the \emph{first regulatory sandbox}, built by running AI workloads inside TEEs. We demonstrate \method using agentic workloads.
For compliance, we derive an \emph{attested policy manifest} (with the applicable regulations, the expected tool calls, and the permitted system calls) for a given user prompt. 
Before an \emph{executor} performs an action, an \emph{attested reference monitor} compares it against the attested policy manifest and allows it only if it matches. Every action and verdict is recorded in an \emph{attested execution graph}, enabling regulators to verify the model's execution and its compliance.
We show that \method meets all requirements, and can check for compliance with regulations and deviations from intended behavior (e.g., due to prompt injection); and incurs low overhead ($<$1\%).
\begin{arxiv}
\footnotetext[1]{Author completed this work at University of Waterloo and is now employed at MIT Lincoln Laboratory.}
\end{arxiv}
\end{abstract}

\section{Introduction}\label{sec:introduction}

Model developers (e.g., Anthropic and OpenAI) are designing and deploying increasingly capable AI systems, including large generative models used as chatbots and agents. These raise concerns about security, privacy, transparency, and safety, highlighting the need for stronger governance through transparent design and evaluation of their model behavior.
Various jurisdictions have already introduced, or are developing AI regulations which will take effect in the near future~\cite{iappjurisdictions,mideastuae,aisafetysandbox,fpfsandboxes,iappsingapore,ibasandboxes,whitecasespain,euaiactsandbox_overview,euaiactarticle57}. 
A key requirement is the establishment of \emph{AI regulatory sandboxes} as
``frameworks for testing AI systems in controlled environments that [...] facilitate development, training, testing, and validation before market entry''~\cite{aisafetysandbox,fpfsandboxes,iappsingapore,ibasandboxes,whitecasespain,euaiactsandbox_overview,euaiactarticle57}.
The current regulations do not spell out the technical mechanisms to realize them.
Using the EU's AI Act~\cite{euaiactarticle57,euaiactsandbox_overview}, we identify the design requirements for regulatory sandboxes:
\begin{itemize}[leftmargin=*,topsep=0pt,itemsep=0pt]
\item \textbf{Compliance:} ``Sandboxes shall [...] improve legal certainty to achieve regulatory compliance with this regulation or other applicable [...] laws'' and ``ensure that appropriate safeguards are in place''.
Thus, the model behavior should comply with relevant laws, while resisting deviations from intended behavior, i.e., unauthorized tool and system calls induced by prompt injection attacks~\cite{usenixBenchmark,zhan2024injecagent,debenedetti2024agentdojo,greshake2023not}. 
The OpenAI–Hugging Face incident illustrates both unauthorized system calls and tool calls by agents~\cite{openai2026huggingface}.
\item \textbf{Confidentiality:} The sandbox should protect ``intellectual property rights and confidential business information or trade secrets [...] including source code'' and requires ``adequate and effective cybersecurity measures to protect the security and confidentiality of the information and data obtained'' [Article 78]. This will incentivize companies to participate in the sandbox and share their models.
\item \textbf{Attestability:} ``The competent authority shall provide a written proof of the activities carried out in the sandbox [and] an exit report detailing [them] and the results'' [Article 57].
This suggests generating tamper-resistant evidence of model behavior for monitoring and verification. Further, ``sandboxes shall be designed and implemented [to] facilitate cross-border cooperation between national competent authorities'' and ``with the competent authorities of other Member States'' [Article 57]. 
This requires that the generated evidence can be checked independently by multiple regulators, which we refer to as \emph{verifiers} hereafter.
\item \textbf{Scalability:} Although the EU AI Act and other related regulations treat interconnected or chained AI architectures (e.g., AI Agents) as a single system~\cite{eu_ai_act_multi_agent}, regulatory sandboxes must scale to handle both single- and multi-agent systems with single- or multi-task objectives.
\end{itemize}
While prior works have discussed regulatory sandboxes conceptually~\cite{waseem2026engineering,ahern2025operationalising,buscemi2025sandboxes,piao2025agentbay}, their concrete design and implementation are missing.
Our goal is to address this gap and develop technical mechanisms to realize regulatory sandboxes.

Several related works, including information-flow control~\cite{costa2025securing,cai2026ghost,siddiqui2025permissive,TiwariIFC,wu2024isolategpt,wang2025agentspec,chen2025shieldagent,hua2024trustagent,jiang2026chaincaps,abaev2026agentguardian,shi2025progent}, prompt-injection defenses~\cite{alam2026prompt,Wang2026AligningPW,kolluri2026optimizing,kim2025prompt,li2026drift,zhong2025rtbas}, contextual integrity~\cite{ngong2026agentscope,tsai2025contextual,bagdasarian2024airgapagent,ghalebikesabi2025privacy}, and sandboxes~\cite{sidik2026beyond,meng2025cellmate,nanda2026nest,googlecloud2026agentsandbox,azure_sandbox,shu2016study,minibox,liang2003isolated}, provide isolated execution and address specific compliance aspects. However, none cover all compliance requirements, including regulatory checks and detection of both prompt injections and unauthorized system calls. Furthermore, they do not provide confidentiality, attestability, and scalability. Trusted execution environments (TEEs) provide hardware-isolated confidentiality and integrity with attestability to multiple verifiers, but require additional mechanisms for compliance.

We present \method{}, the first TEE-based regulatory sandbox designed to meet all the above requirements. 
We demonstrate it for agentic workloads where a \emph{planner} generates an ordered sequence of expected tool calls for an \emph{executor} to follow~\cite{piet2026web,del2025architecting,beurer2025design}. 
Before execution, \method{} converts each user prompt into an \emph{attested policy manifest} (APM) specifying the applicable regulations, the expected tool-call sequence, and the permitted tools and system calls under least privilege, and any stop conditions (e.g., timeout). 
An \emph{attested reference monitor} (\RefMon) checks each executor action against the APM, and blocks any violations. 
\method{} records actions and \RefMon decisions in an \emph{attested execution graph} (AEG). 
\method{} uses the TEE's remote attestation to construct verifiable evidence of the APM, AEG, and \RefMon decisions, enabling multiple verifiers to check them independently.
We prototype and evaluate \method on Intel TDX and NVIDIA H100 GPU across realistic agentic workloads involving interrupts, single- and multi-task, and single- and multi-agent settings.
We claim the following contributions:
\begin{itemize}[leftmargin=*,itemsep=0pt,topsep=0pt]
\item identify the design requirements for a regulatory sandbox from the EU's AI Act, covering compliance, confidentiality, attestability, and scalability; (\S\ref{sec:problem})
\item present \method\footnote{Code will be publicly available upon publication.}, the \emph{first regulatory sandbox} designed to meet the various design requirements by running the models inside TEEs; (\S\ref{sec:approach})
\item evaluate \method across realistic agentic workloads showing that it correctly checks regulatory compliance and detects deviations from intended behavior, i.e., unauthorized tool and system calls (e.g., from prompt injection); and incurs low overhead for attestations ($<$1\%). (\S\ref{sec:setup} and \S\ref{sec:evaluation})
\end{itemize}
\section{Background}\label{sec:background}

\noindent\textbf{Agentic systems} support multi-modal inputs (e.g., text, images), with models fine-tuned to generate plans given user prompts, and invoke tools and APIs~\cite{Wei2026AgenticRF}. These workloads are prevalent where even chatbot services (e.g., ChatGPT and Claude) support tool-use beyond simple conversations.
Typically, a single agent both plans and executes a user-requested task. 
However, if compromised, the agent can manipulate subsequent plans to execute malicious instructions. Recent work therefore advocates a dedicated \emph{planner} that generates a plan, which is executed by an \emph{executor}~\cite{piet2026web,del2025architecting,beurer2025design}.
In a \emph{single-agent} setting, a user's prompt may comprise either a single task (``single task-single agent'' or ST-SA) or execute multiple tasks (``multiple tasks-single agent'' or MT-SA). 
In case of multiple tasks, the agent can execute them sequentially or interleave them (initiate another task while waiting for the result of the first). Such interleaving can result in complex control flows.
In case of multiple tasks, an agent can distribute them among several sub-agents, which independently execute their assigned tasks and return the results (``multiple tasks-multiple agents'' or MT-MA).
Here, each sub-agent may execute a single sub-task (if the number of tasks is the same as the number of sub-agents), or multiple sub-tasks (if the number of sub-agents is less than the number of sub-tasks).


\noindent\textbf{Regulations on AI sandboxes} include the European Union's AI Act where Article~57 requires member states to establish at least one national AI regulatory sandbox for developing, testing, and validating AI systems before market release~\cite{euaiactarticle57}. 
Spain's Royal Decree 817/2023 established a regulatory sandbox for high-risk systems~\cite{whitecasespain}.
Various other countries, such as the United Kingdom~\cite{ibasandboxes}, Singapore~\cite{fpfsandboxes}, Brazil, the United States~\cite{fpfsandboxes,aisafetysandbox}, Japan and South Korea~\cite{aisafetysandbox}, and the United Arab Emirates~\cite{mideastuae}, have all explored sandbox initiatives for AI, or expanded their existing regulatory sandboxes to include AI.


\noindent\textbf{Trusted execution environments (TEEs)} provide hardware-based isolation of code and data, ensuring confidentiality and integrity against a potentially compromised host system.
TEEs also support \emph{remote attestation}, where a TEE generates an attestation report (e.g., an Intel TDX \textit{quote}) containing a signed measurement of its loaded code and initial state.
A remote verifier can validate the signature against the manufacturer's certificate chain, and compare the measurement against an expected value to confirm the integrity of the loaded code and data.
Multiple independent verifiers can check the same report without trusting each other or the \textit{prover}.
Intel TDX~\cite{tdx} is one TEE architecture that realizes \textit{confidential virtual machines} (CVMs) in the form of \textit{trust domains} (TDs).
For AI workloads, proprietary models can execute inside TEEs without exposing model weights, prompts, or intermediate outputs to untrusted sources. 
Recent advances such as NVIDIA's H100 GPU~\cite{h100} extend confidential computing to GPUs for generative models.


\noindent\textbf{Control flow integrity (CFI)}~\cite{hek-cfi,typesqueezer,flowguard,sherloc,kuzhiyelil2020towards,pacstack,pal,pac-it-up} prevents a program from deviating from its intended control flow due to memory corruption vulnerabilities (e.g., buffer overflows or use-after-free memory vulnerabilities). 
CFI restricts indirect control flow transfers to a set of destination addresses that are valid according to a particular policy, which specifies which control-flow transitions are permitted~\cite{sok_cfa_cfi}.
%
For AI agents, deviations from intended behavior need not result from corruption of a return address or function pointer like classical CFI. 
For example, prompt injection may cause an agent to invoke a tool that is not part of its intended execution path~\cite{greshake2023not,usenixBenchmark,zhan2024injecagent,debenedetti2024agentdojo}. This is analogous to a control-flow violation though the underlying cause and mechanism differ.

Some attacks can alter program state without necessarily violating a control-flow policy. Thus,
\noindent\textbf{data flow integrity (DFI)}~\cite{castro2006securing,song2016enforcing,buccafurri2024mqtt,chen2022finedift,hdfi,datashield} complements CFI by constraining how data propagates through a program. Rather than restricting \emph{where execution can go}, DFI restricts \emph{which inputs can influence data}.  A DFI policy includes permitted data dependencies: a value originating from an untrusted or otherwise restricted \emph{source} may be prevented from influencing a sensitive \emph{sink} unless authorized. 
%
For AI agents, analogous concerns arise when the tool calls are executed as intended but the data and inputs passed to the tools differ.
We summarize the frequently used notations and their descriptions in Table~\ref{tab:notation}.

\begin{table}[!h]
\centering
\footnotesize
\setlength{\tabcolsep}{0pt}
\renewcommand{\arraystretch}{1.15}
\caption{\textbf{Summary of Notations}}
\label{tab:notation}
\resizebox{0.85\columnwidth}{!}{%
\begin{tabular}{@{}l@{\hspace{7pt}}p{0.78\columnwidth}@{}}
\toprule
\textbf{Notation} & \textbf{Description} \\
\midrule
$p$ & User prompt \\
$C$ & System-call manifest of the authorized tools \\
$R$ & Applicable regulations, $R \in \{\text{GDPR},\ \text{HIPAA}\}$ \\
$\mathrm{stop}$ & External stop condition (e.g., maximum attempts) \\
\midrule
$\traj$ & Expected trajectory, $\traj=\langle e_1,\dots,e_n\rangle$ \\
$e_j$ & Expected tool call at position $j$ of $\traj$ \\
$\Gamma$ & Set of guards in the APM \\
$\Gamma_{\mathrm{reg}}$ & Regulation guard \\
$\Gamma_{\mathrm{traj}}$ & Trajectory guard \\
$\Gamma_{\mathrm{sys}}$ & System call guard \\
$\Gamma_{\mathrm{int}}$ & Stop guard \\
$\Gamma_{\mid \act}$ & Guards covering the action $\act$ \\
$\mathrm{Auth}$ & Authorized system calls, paths, and limits \\
\midrule
$\act$ & A single executor tool call that \RefMon mediates \\
$\varphi$ & Five-element action tuple derived from $\act$ \\
$J$ & Fixed judge model that evaluates $\Gamma_{\mathrm{reg}}$ \\
$\sigma$ & Run state tested by $\Gamma_{\mathrm{int}}$ \\
$v$ & \RefMon's verdict: \textsc{allow} or \textsc{block} \\
\midrule
$\mathrm{H}(\cdot)$ & Cryptographic hash \\
$m_i$ & Measurement of AEG node $i$ \\
$\mathit{seq}_i$ & Commit sequence number of AEG node $i$ \\
$c_i$, $c_N$ & Hash-chain value at node $i$, and the chain head \\
$s$ & Structure root over the AEG topology \\
$w$ & Model weights, bound as \texttt{weights\_hash} \\
$\signatt$ & Attestation signature produced in the TD \\
\midrule
$t$, $m$ & Number of tasks and executor sub-agents \\
$\adv$ & Adversary \\
\bottomrule
\end{tabular}%
}
\end{table}

\section{Problem Statement}\label{sec:problem}

Our goal is to design a regulatory sandbox by identifying key design requirements from existing regulations, and by examining how existing work falls short.

\noindent\textbf{\underline{Setting:}} A user provides a textual prompt describing a task that requires invoking tools from multiple toolkits (e.g., calls to remote applications via APIs like Gmail or Amazon, and even local tools like Linux commands or system calls).
During execution, the agent accesses untrusted external sources and may invoke multiple tools. For example, a user may ask the agent to retrieve product reviews from a website, summarize them, and email the results. The agent would invoke \texttt{GetProductReviews} to access the reviews, summarize the results, and then invoke \texttt{SendEmail} to share them with the specified recipient. 
The user can interrupt execution at any time to stop the ongoing operation, or there can be other stop conditions such as timeouts.
We consider multiple verifiers who seek to check the agent's behavior against the applicable regulations, and to provide guidelines for ensuring compliance.

\noindent\textbf{\underline{Threat and System Model:}} We assume an adversary \adv with knowledge of the available tools and their corresponding function calls. \adv can manipulate only external sources accessible to the executor agent to force the agent to deviate from its intended behavior by executing malicious instructions. 
Such deviations take the form of unauthorized tool calls, and unauthorized system calls to access system resources, data exfiltration, or privilege-escalation attacks.
This threat model follows prompt injection attacks~\cite{greshake2023not,usenixBenchmark,zhan2024injecagent,debenedetti2024agentdojo}.
Furthermore, the planner and executor are standard open-weight models obtained from trusted public sources.

The server running the agent is untrusted (including the host operating system and hypervisor), but it is equipped with CVM and GPU TEEs. 
The control logic of the agent runs inside the CVM while the required LLM inference operations execute in the GPU TEE. 
We use Intel TDX's TDs as a CVM and NVIDIA H100 for a GPU TEE.
Verifiers (e.g., users or regulators) accept the TEE's hardware manufacturer (e.g., Intel or NVIDIA), and any attestation that is digitally signed using their secret attestation key is trusted. Attacks that rely on physical access to the machine are out of scope.

\noindent\textbf{\underline{Design Requirements and Challenges:}} Using the EU's AI Act as an example~\cite{euaiactarticle57,euaiactsandbox_overview}, we identify various design requirements for a regulatory sandbox which should show:
\begin{enumerate}[nosep, leftmargin=*,label={\textbf{R\arabic*}}]
\item \label{compliance}\emph{compliance} (with regulations and deviations from intended behavior, i.e., unauthorized tool and system calls);
\item \label{confidentiality}\emph{confidentiality} (protects model weights from verifiers);
\item \label{attestable}\emph{attestability} (produces tamper-resistant evidence of model behavior for monitoring); and
\item \label{scalable}\emph{scalability} (handles single- and multi-task operations with single- and multi-agents involved).
\end{enumerate}
Designing mechanisms that satisfy all requirements is challenging. We identify various challenges in the design of a regulatory sandbox. We need:
\begin{enumerate}[nosep, leftmargin=*, label={\textbf{C\arabic*}}]
\item \label{c1:isolated} \emph{isolated execution environment} which can efficiently execute agentic workloads while meeting (\ref{confidentiality}-\ref{scalable});
\item \label{c2:apm} \emph{policy representation} which is a concise, unified representation of regulations and permitted tool and system calls, tailored to each prompt (\emph{principle of least privilege}), and serves as ground truth for agent behavior;
\item \label{c4:regulation} \emph{regulatory compliance} with mechanisms to check agent behavior against semantic policy requirements, requiring reasoning over regulation text rather than direct check against fixed policies;
\item \label{c3:cfi} \emph{execution-flow enforcement} using efficient and attestable mechanisms to check control and data flows and intervene before unauthorized actions occur; and
\item \label{c5:aeg} \emph{execution verification} using a compact representation of complex control and data flows across diverse single- and multi-agent workloads (including interleaving control flows) that is easy for verifiers to check.
\end{enumerate}

\begin{table*}[t]
\centering
\footnotesize
\setlength{\tabcolsep}{5pt}
\renewcommand{\arraystretch}{1.05}
\caption{\textbf{Overview of Literature:} \cmark (requirement supported); \xmark (requirement not supported). \emph{Compliance} (\ref{compliance})covers regulations, and deviations from intended behavior via unauthorized tool or system calls. No prior work meets all the requirements.}
\label{tab:related}
\resizebox{0.85\textwidth}{!}{%
\begin{tabular}{@{}l ccc c c c@{}}
\bottomrule

\toprule
& \multicolumn{3}{c}{\textbf{Compliance} (\ref{compliance})} & \textbf{Confidentiality} & \textbf{Attestability} & \textbf{Scalability} \\
\cmidrule(lr){2-4}
& Regulations & Unauth. Tool Calls & Unauth. Sys. Calls & (\ref{confidentiality}) & (\ref{attestable}) & (\ref{scalable}) \\
\bottomrule

\toprule
Regulatory Sandboxes~\cite{waseem2026engineering,ahern2025operationalising,buscemi2025sandboxes} & \xmark & \xmark & \xmark & \xmark & \xmark & \xmark \\
Agent Sandboxes~\cite{sidik2026beyond,meng2025cellmate,nanda2026nest,googlecloud2026agentsandbox} & \xmark & \cmark & \xmark & \xmark & \xmark & \xmark \\
Traditional Sandboxes~\cite{azure_sandbox,shu2016study,minibox,liang2003isolated} & \xmark & \xmark & \cmark & \xmark & \xmark & \xmark \\
Information-flow Control~\cite{costa2025securing,cai2026ghost,siddiqui2025permissive,TiwariIFC,wu2024isolategpt,wang2025agentspec,chen2025shieldagent,hua2024trustagent,jiang2026chaincaps,abaev2026agentguardian,shi2025progent} & \xmark & \cmark & \cmark & \xmark & \xmark & \cmark \\
Contextual Integrity~\cite{ngong2026agentscope,tsai2025contextual,bagdasarian2024airgapagent,ghalebikesabi2025privacy,li-contextlens,fan2024goldcoin,li2025privacy} & \cmark & \xmark & \xmark & \xmark & \xmark & \cmark \\
Control-flow Integrity~\cite{debenedetti2025defeating,foerster2026camels,jha2026breaking,wang2025agentarmor,alam2026prompt,Wang2026AligningPW,kolluri2026optimizing,kim2025prompt,li2026drift,zhong2025rtbas} & \xmark & \cmark & \xmark & \xmark & \xmark & \cmark \\
Formally Verified Agents~\cite{balunovic2024ai,palumbo2026formal,solko2026efficient,kravchenko2026agentic} & \xmark & \cmark & \xmark & \xmark & \xmark & \cmark \\
Agents in TEEs~\cite{abdollahi2026agentee,villa2026prismata,bodea2025trusted,duddu2024laminator,grigor2026vet,chantasantitam2026pal} & \xmark & \xmark & \xmark & \cmark & \cmark & \xmark \\
\midrule
\textbf{\method{} (this work)} & \cmark & \cmark & \cmark & \cmark & \cmark & \cmark \\
\bottomrule

\toprule
\end{tabular}%
}
\end{table*}

\noindent\textbf{\underline{Limitations of Existing Work:}} Existing research has explored isolated execution for agents, thus, meeting some compliance requirements (e.g., deviation due to prompt injections); or running agents inside TEEs for either confidentiality or attestability. But none of these approaches meet all of the aforementioned requirements for regulatory sandboxes.
Table~\ref{tab:related} summarizes prior work against our design requirements, and we discuss them below:
\begin{itemize}[leftmargin=*,itemsep=0pt,topsep=0pt]
\item \textbf{Sandboxes:} Existing work on regulatory sandboxes focuses on high-level policy discussions, but lack a concrete implementation~\cite{waseem2026engineering,ahern2025operationalising,buscemi2025sandboxes}.
\emph{Agent sandboxes} ensure that the agent's tool calls does not deviate from intended behavior by checking against static policies~\cite{sidik2026beyond,meng2025cellmate,nanda2026nest,googlecloud2026agentsandbox}, but do not check for regulatory compliance and unauthorized system calls. They also do not meet confidentiality, attestability, and scalability requirements (\ref{compliance}: Regulations, System Calls; \ref{confidentiality}-\ref{scalable}$\rightarrow$\xmark).
\emph{Traditional sandboxes} confine untrusted code at the OS level preventing unauthorized system calls~\cite{azure_sandbox,shu2016study,minibox,liang2003isolated}. 
However, they do not meet other requirements (\ref{compliance}: Regulations, Tool Calls; \ref{confidentiality}-\ref{scalable}$\rightarrow$\xmark).

\item \textbf{Compliance Checking:}
\emph{Contextual integrity} checks whether information flows from agents violate privacy by comparing against privacy norms~\cite{ngong2026agentscope,tsai2025contextual,bagdasarian2024airgapagent,ghalebikesabi2025privacy}.
Some works use the contextual integrity framework to evaluate against legal statutes~\cite{li-contextlens,fan2024goldcoin,li2025privacy} (\ref{compliance}: Regulations $\rightarrow$ \cmark).
\emph{Information-flow control} includes taint tracking or checking confidentiality/integrity labels~\cite{costa2025securing,cai2026ghost,siddiqui2025permissive,TiwariIFC,wu2024isolategpt}, and check actions against policies before granting permission~\cite{wang2025agentspec,chen2025shieldagent,hua2024trustagent,jiang2026chaincaps,abaev2026agentguardian,shi2025progent}.
\emph{Control-flow integrity} mechanisms~\cite{debenedetti2025defeating,foerster2026camels,jha2026breaking,wang2025agentarmor} and some \emph{prompt injection defenses}~\cite{alam2026prompt,Wang2026AligningPW,kolluri2026optimizing,kim2025prompt,li2026drift,zhong2025rtbas} compare the agent's actual behavior against a control- or data-flow graph at the granularity of tool calls.
While these cover deviations from intended behavior, i.e., unauthorized tool and system calls due to prompt injection (across both single-/multi-task and single-/multi-agent), aside from the contextual integrity work, neither of them cover regulatory compliance (\ref{compliance}: Regulations $\rightarrow$\xmark).
Furthermore, none of them meet confidentiality and attestability (\ref{confidentiality}-\ref{attestable} $\rightarrow$\xmark; \ref{scalable}: $\rightarrow$ \cmark): a verifier would have to trust the model developer's report and claims which is not feasible if they are malicious.
Finally, while prior work covers control-flow violations, it overlooks data-flow integrity violations.

\item \textbf{Formally Verified Agents:} These include static guarantees for checking if agent behavior matches intended behavior~\cite{balunovic2024ai,palumbo2026formal,solko2026efficient,kravchenko2026agentic}. 
This requires trusting the model developer for the formal proofs, which is not our threat model (malicious developer). Some works have covered multi-task and multi-agent systems (\ref{scalable}$\rightarrow$ \cmark), but do not support confidentiality (\ref{confidentiality}$\rightarrow$ \xmark), nor do they provide the tamper-resistant evidence required for attestability (\ref{attestable} $\rightarrow$ \xmark).

\item \textbf{Agents in TEEs:}
Prior works have used TEEs to provide confidentiality~\cite{abdollahi2026agentee,villa2026prismata,bodea2025trusted} and remotely-verifiable evidence~\cite{duddu2024laminator,grigor2026vet,chantasantitam2026pal} of agentic workloads executing within them for attestability and scalability.
However, scaling them for agentic workloads is an open problem where additional mechanisms are required to check an agent's behavior for compliance (\ref{compliance} $\rightarrow$ \xmark).
\end{itemize}
\begin{takeaway}
\textbf{Takeaway:} Designing a regulatory sandbox that satisfies all requirements (\ref{compliance}-\ref{scalable}) is an open problem. None of the prior solutions can be directly applied to meet all the challenges (\ref{c1:isolated}-\ref{c5:aeg}), thus, requiring new mechanisms.
\end{takeaway}
\section{Design of \method}\label{sec:approach}

\begin{figure}[htb]
    \centering
    \includegraphics[width=\columnwidth]{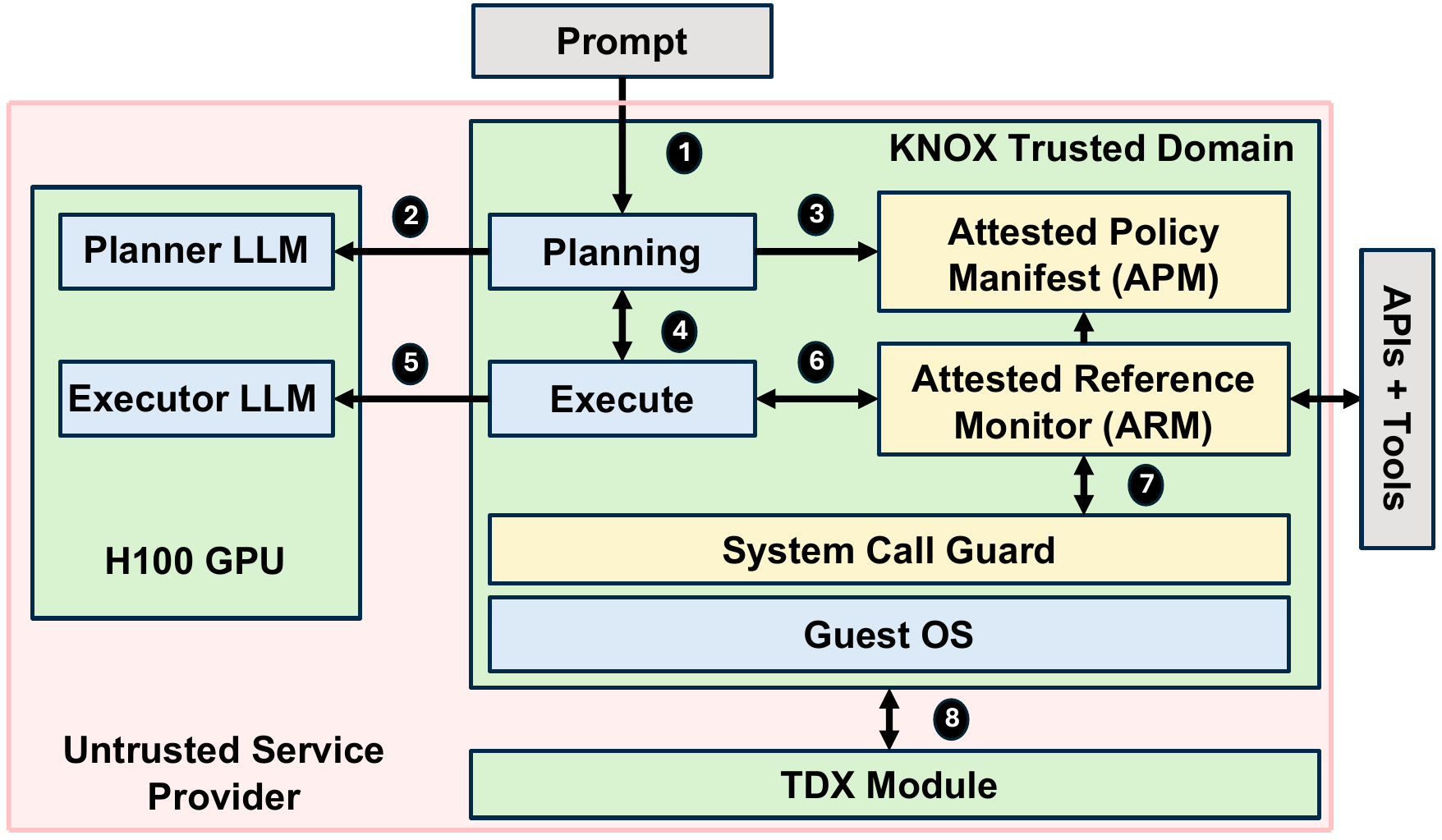}
    \caption{\textbf{Overview of \method:} Planner, executor, and \RefMon run inside Intel TDX's trusted domain. The \emph{planner} derives the expected trajectory from the clean prompt and synthesizes the APM tailored to it, which \RefMon enforces on every executor action. \RefMon mediates every executor action crossing the trust boundary, checking for deviation due to prompt injection, and compliance with regulations, along with \emph{OS-level sandboxing} that confines unauthorized executor system calls.}
    \label{fig:overview}
\end{figure}

We present \method, the \emph{first hardware-assisted regulatory sandbox} that meets all the aforementioned requirements, as highlighted below (Figure~\ref{fig:overview}):
\begin{itemize}[leftmargin=*,itemsep=0pt,topsep=0pt]
\item \textbf{Isolated Execution (\ref{c1:isolated}):} 
\method{} leverages TEEs for isolated execution, confidentiality, and attestability (\ref{confidentiality}--\ref{scalable})~\cite{duddu2024laminator,chantasantitam2026pal}.

\item \textbf{Policy Representation (\ref{c2:apm}):} We introduce an \emph{attested policy manifest} (APM) that unifies regulations, permitted tool and system calls for each prompt, enforcing least privilege by including only task-relevant permissions (\S\ref{sec:apm}).

\item \textbf{Regulatory Compliance (\ref{c4:regulation}):} We draw on contextual integrity to represent the executor's actions defined by sender, receiver, subject, data type, and transmission principle, and use a language model as a judge to reason about their compliance with regulations (\S\ref{sec:apm}).

\item \textbf{Execution-flow Enforcement (\ref{c3:cfi}):} We introduce an \emph{attested reference monitor} (\RefMon) that checks each action against the APM, and blocks any action the APM does not permit (\S\ref{sec:refmonitor}).

\item \textbf{Execution Verification (\ref{c5:aeg}):} We introduce an \emph{attested execution graph} (AEG) that (a) captures control and data flows and (b) records whether each action was permitted or blocked by \RefMon (\S\ref{sec:aeg}). This enables verifiers to check the executor's behavior against the APM. We define the various attestations in Appendix~\ref{app:attestations}.
\end{itemize}

\noindent\textbf{\underline{Terminology:}} We use the following terms:
\begin{itemize}[leftmargin=*,topsep=0pt,itemsep=0pt]
    \item an \emph{action} is a single executor tool call, the unit \RefMon mediates and attests (one \texttt{tool\_call} node in the AEG).
    \item a \emph{flow} is a single information-flow test case. For $\Gamma_{\mathrm{reg}}$, a flow is a read-then-disclose pair of actions, while for $\Gamma_{\mathrm{traj}}$, $\Gamma_{\mathrm{sys}}$, and $\Gamma_{\mathrm{int}}$ it is realized by a single action.
    \item a \emph{bundle} groups $t$ flows (one per task) into a single task-agent prompt executed over $m$ sub-agents. A bundle describes the unit executed by a workload.
    \item a \emph{toolkit} is an application or service, such as Gmail or Slack, that provides multiple tools for specific actions, such as \texttt{SendMail} in Gmail (\S\ref{sec:setup}).
\end{itemize}

\subsection{Attested Policy Manifest}\label{sec:apm}

For each prompt, we generate an \emph{attested policy manifest} (APM) once, which captures the expected behavior of the executor against which its actual behavior is compared. 
The planner synthesizes the APM once per prompt: $(p,C,R,\mathrm{stop}) \longmapsto \text{APM}=(\traj,\Gamma)$\footnote{\textbf{Notation:} We use $\Gamma$ for a set of guards whereas we use $\Gamma_{<.>}$ to a single guard specific to regulations, tool calls, system calls, or interrupts.}, where $p$ is the prompt, $\traj$ is the expected trajectory, and $\Gamma$ is the set of guards corresponding to the system-call manifest $C$ (permitted tools, parameter schemas, and required system calls), applicable regulations $R$, and the external $\mathrm{stop}$ condition (e.g., the maximum number of attempts).
Each guard states \emph{what} is permitted, and is the ground truth to check the executor's actual behavior.
The \emph{trajectory} $\traj \;=\; \langle e_1,\dots,e_n\rangle$ is the ordered tuple of expected
tool calls from the planner. Each expected tool call $e_j$ has a \emph{trajectory guard} $\Gamma_{\mathrm{traj}}$ that permits action $\act$ only if its tool and parameters match $e_j$ at position $j$, verified by comparing their hashes: $\mathrm{H}(\act)=\mathrm{H}(e_j)$.



\noindent\textbf{\underline{Types of Guards:}} We identify the various types of guards:
\begin{itemize}[leftmargin=*,topsep=0pt,itemsep=0pt]
\item \textbf{Regulation Guard ($\Gamma_{\mathrm{reg}}$):}
The APM includes the applicable regulations $R \in \{\text{GDPR}, \text{HIPAA}\}$ which are analyzed to check whether an executor's action is permitted or not. Each executor action is separated into a five-element \emph{action tuple} (\textit{sender}, \textit{recipient}, \textit{subject}, \textit{information\_type}, \textit{transmission\_principle}), given by $\varphi=\derive(\act)$.
$\Gamma_{\mathrm{reg}}$ is the ``ground truth'' of \emph{which} law governs which actions, not the compliance verdict.
For example, assume a hospital's scheduling agent sends a patient's lab results to an insurer. Then, $\varphi=$(hospital agent, insurer, patient, lab results, payment). The sender and information type indicate HIPAA's Privacy Rule applies, and the transmission principle (payment) determines which provision is relevant. If only the recipient changes to a marketing vendor, the applicable provision and outcome differ, even though the other four elements are unchanged.



\item \textbf{Trajectory Guard ($\Gamma_{\mathrm{traj}}$):} The planner emits the ordered expected tool calls $e_1,\dots,e_n$ that form $\traj$, each stored with its hash $\mathrm{H}(e_j)$. $\Gamma_{\mathrm{traj}}$ is static and exact: a proposed action is on-trajectory if it equals the expected tool call at the current position, $\mathrm{H}(\act)=\mathrm{H}(e_j)$. This catches both deviation from intended tool usage and system calls.

\item \textbf{System Call Guard ($\Gamma_{\mathrm{sys}}$):}
A compromised executor can issue system calls directly from inside a tool's execution, where \RefMon has no visibility. 
$\Gamma_{\mathrm{sys}}$ checks against such unauthorized system calls.
Rather than a single global blocklist of system calls applied to every prompt, the planner derives the authorized set given the tools that a prompt's trajectory actually uses. 
Each toolkit (e.g., Gmail) lists the system calls, filesystem paths, and resource limits that its tools (e.g., \texttt{SendMail}, \texttt{ReadMail}) need. 
For a given prompt, the planner generates an authorized set $\mathrm{Auth}$ by combining system calls required for each toolkit being used and for model runtime (e.g., \texttt{libc}, \texttt{/proc}).
An action is allowed only if every system call the executor issues is in $\mathrm{Auth}$. 
For instance, a prompt whose trajectory $\traj$ never opens a terminal cannot call \texttt{execve} or \texttt{socket}.

\method{} enforces $\mathrm{Auth}$ with three kernel mechanisms, applied to the executor process before
any tool runs:
\begin{itemize}[leftmargin=*,topsep=0pt,itemsep=0pt]
\item \emph{seccomp (which system call?):} seccomp filters every system call at entry by its
number, returning \texttt{EPERM} for anything outside $\mathrm{Auth}.\mathrm{sys}$: privilege changes
(\texttt{setuid}), mount and namespace manipulation (\texttt{mount}, \texttt{unshare}), kernel-module
loading (\texttt{init\_module}), process tampering (\texttt{ptrace}), and network access or program
execution (\texttt{socket}, \texttt{execve}) are denied unless required. 

\item \emph{Landlock (which file?):} seccomp sees only system call numbers and register values, so it cannot differentiate between \texttt{open("/etc/shadow")} from the executor opening its own file. 
Landlock, an unprivileged Linux security module, adds path-based control, confining file operations (read, write, create, remove) to $\mathrm{Auth}.\mathrm{paths}$.

\item \emph{rlimits (how much?):} Some system calls are dangerous only in aggregate. \texttt{fork} is needed for normal operation, so banning it breaks the runtime; the threat is \emph{unbounded} creation, which seccomp cannot detect since it is stateless. POSIX resource limits cap the totals in $\mathrm{Auth}.\mathrm{res}$: \texttt{RLIMIT\_NPROC} caps the process count (a fork bomb fails while ordinary forking succeeds) and \texttt{RLIMIT\_FSIZE} bounds file growth.
\end{itemize}
All three are complementary, enforcing different kernel-layer decisions that the others cannot. 
Together, they mediate the entire in-tool system call surface, complementing \RefMon's mediation, and providing \emph{defense-in-depth}. 
As $\mathrm{Auth}$ is derived per prompt, $\Gamma_{\mathrm{sys}}$ also prevents \emph{confused-deputy via an unused system call}, where a tool invokes a system call required by another toolkit but unnecessary for the current prompt.
The TDX quote attests to TD's launch state, measuring the guest kernel and \texttt{initrd} through the build-time measurement register (MRTD) and the runtime measurement registers (RTMRs). However, it does not directly attest to per-prompt seccomp-BPF, Landlock, or rlimit policies, which are bound only indirectly through the APM measurement and its permitted set $\mathrm{Auth}$. Thus, the quote alone cannot verify that the derived filters were installed. Extending an RTMR with the applied BPF program and Landlock ruleset would close this gap and is a concrete hardening step. This is left as future work.
\item \textbf{Stop Guard ($\Gamma_{\mathrm{int}}$):}
Given various stop conditions and the resource limitations described in the system-call manifest $C$, the planner attaches a static predicate on the system state: execution is stopped whenever a timeout, resource cap, or external stop signal is present. 

\end{itemize}
Appendix~\ref{app:example}: Figure~\ref{fig:apm-guards} shows the guards for an example trajectory that uses shell and e-mail tools.

\noindent\textbf{\underline{Attesting Policy Manifest:}} We attest the APM generation process so that verifiers can check it. The prover (i.e., model developer) runs the APM generation inside the TEE and outputs an attestation binding the generated APM, containing $\traj$ and $\Gamma$, to the inputs ($p$, $C$, $R$, $\mathrm{stop}$). Since the attestation is signed using the trusted TEE manufacturer's attestation key (e.g., Intel or NVIDIA), a verifier can check the signature against the hardware manufacturer's certificate chain and learn that the assertion holds. 
A verifier confirms that the enforced policy was legitimately derived from the user's clean prompt and was neither replaced nor tampered with. Because $\mathrm{Auth}$ is signed, each run also provides a verifiable \emph{system-call receipt}, allowing verifiers to recompute exactly which system calls, paths, and resources the prompt authorized.
The attestation is defined in Appendix~\ref{app:attestations}: \Cref{att:apm}.

\subsection{Attested Reference Monitor}\label{sec:refmonitor}

Every interaction the executor has with the outside world is mediated by \RefMon, which checks each proposed action against APM. \RefMon ensures complete mediation, i.e., no tool executes except through it, and it runs inside the TEE and every decision is attested (Appendix~\ref{app:attestations}: \Cref{att:check}).
Specifically, an \emph{executor} runs the agentic loop and proposes the next tool call each turn; and only the executor is exposed to (possibly prompt injected) tool outputs. 
Each proposed action $\act$, together with its output, passes through \RefMon, which \emph{admits $\act$ iff it matches the expected tool call at its position in the APM's trajectory $\traj$ and satisfies every guard in $\Gamma$ applicable to it}.
The APM states \emph{what} is permitted, while \RefMon decides \emph{how} each guard is applied. 
Each guard is discharged in one of two ways, either by a \textsc{static} predicate or by an \textsc{oracle}, and \RefMon records which one it used.
\begin{itemize}[leftmargin=*,topsep=0pt,itemsep=0pt]
\item \textsc{static} is an exact predicate over $\act$ and APM: $\Gamma_{\mathrm{traj}}$ checks $\mathrm{H}(\act)=\mathrm{H}(e_j)$; $\Gamma_{\mathrm{sys}}$ checks
that every system call an action issues is in $\mathrm{Auth}$; $\Gamma_{\mathrm{int}}$ tests the trigger set over the run state $\sigma$. 
These are sound by construction as they admit no violating action and never block a permitted one, which our evaluation confirms empirically (\S\ref{sec:evaluation}). 

\item \textsc{oracle} is a language model judge $J$ over $\varphi=\derive(\act)$ which identifies the action tuple for $\act$.
Only $\Gamma_{\mathrm{reg}}$ is evaluated this way, by semantic reasoning of whether $\varphi$ is permitted under the applicable regulations $R$. Since $J$ is a non-deterministic language model, its accuracy is \emph{empirically evaluated} and not proven.
\end{itemize}


\noindent\textbf{\underline{Regulatory Compliance:}} Given the regulation text $R \in \{\text{GDPR},\ \text{HIPAA}\}$, \RefMon queries a \emph{fixed} judge model $J$ (with chain-of-thought and $k$-shot examples) to classify whether an executor's action is compliant with $R$: \{\texttt{compliant}, \texttt{non-compliant}, \texttt{not-applicable}\}. A \texttt{non-compliant} action is blocked and recorded, while \texttt{not-applicable} requires manual verification (Appendix~\ref{app:attestations}: \Cref{att:reg}).

\noindent\textbf{\underline{Trajectory Deviation:}} The ground truth for a legitimate action $\act$ is the trajectory $\traj$ itself (example in Appendix~\ref{app:example}: Figure~\ref{fig:apm-guards}). 
\RefMon tracks the next unmatched expected tool call $e_j$ and admits $\act$ iff it equals $e_j$ in tool identity, parameters, and position, decided by the trajectory guard $\mathrm{H}(\act) = \mathrm{H}(e_j)$.
On a match, $\act$ is legitimate and the pointer advances ($j \!\leftarrow\! j{+}1$); on a mismatch, $\act$ is an off-trajectory deviation and is blocked (the tool is not executed and a \texttt{blocked} node is recorded as discussed in \S\ref{sec:aeg}). 
The check is order-aware, so an out-of-position tool call is blocked. 
This catches both unauthorized tool use (and incorrect parameters) and unauthorized system calls. 

\noindent\textbf{\underline{Runtime Interrupt and Stop Condition:}}
Some executions might be stopped at runtime either by a user's interrupt or because some stop condition is met (e.g., number of failed attempts for some task or other conditions from resource constraints).
This static guard is deterministic and mediates such an action to an \texttt{interrupt} node directly (discussed in \S\ref{sec:aeg}). We consider the following two cases:
\begin{itemize}[leftmargin=*,topsep=0pt,itemsep=0pt]
\item \emph{Process-level interrupt} (crash, OOM/host kill, or operator pause) discards the entire in-memory run, including any partially produced attestations. 
\item \emph{Action-level interrupt} by
$\Gamma_{\mathrm{int}}$ or because an attestation cannot be produced, \RefMon does not execute or commit the action but instead records the stop as an explicit attested interrupt event (Appendix~\ref{app:attestations}: \Cref{att:int}). In multi-agent runs, only the affected sub-agent stops at this point; the others
run to completion.
\end{itemize}

\subsection{Attested Execution Graph}\label{sec:aeg}

To make it easier to follow the control and data flow for verifiers, we propose an \emph{attested execution graph} (AEG) where every action that \RefMon admits or rejects is recorded as a node in an AEG and carries its own evidence.
The AEG is the object a verifier checks: it shows not only \emph{what} the executor did but that each action was produced by the claimed model, checked, and either executed or stopped.
The AEG is a hash chain over committed measurements. 
Each node $i$ carries $m_i=\mathrm{H}(\textit{payload}_i)$ over its canonical content, and the chain folds nodes in commit order:
$$
c_i=\mathrm{H}\bigl(c_{i-1}\,\|\,\textit{seq}_i\,\|\,m_i\bigr),\qquad c_0=\mathrm{H}(\varnothing).
$$
Here, $\textit{seq}_i$ is the node's commit sequence number, a counter \RefMon increments as it appends each node, which binds every measurement to its position in the run and not only to its content.
Thus, any omission, reordering, or modification changes the head $c_N$. 
The head is a purely software value: any verifier recomputes it by walking the chain, but on its own it proves nothing about who produced the AEG, since anyone can compute a consistent head for a forged AEG. 
Since the chain captures only content and order, we additionally compute a \emph{structure root} $s$ over the graph topology, including each node's kind, parent, and dispatched branch set, binding multi-agent bifurcation and recombination. 
The AEG summary is $(c_N,s)$, and a \emph{single} quote per AEG binds it into the quote's \texttt{report\_data}, where $(c_N,s)$ is attested. 
As the head already commits to every node's measurement and the structure root to the topology, this one quote transitively attests the entire AEG (evidence that the TD actually ran this AEG; Appendix~\ref{app:attestations}: \Cref{att:aeg}).

\begin{figure*}[htb]
\centering
\resizebox{\textwidth}{!}{%
\begin{tikzpicture}[
  >={Stealth[round]},
  font=\small,
  node distance=3.4mm and 8mm,
  base/.style={draw, rounded corners=2pt, minimum height=5.2mm, minimum width=14mm,
               align=center, inner sep=1.5pt},
  prompt/.style={base, fill=black!6},
  disp/.style  ={base, fill=orange!12},
  pln/.style   ={base, fill=blue!14},
  t1/.style    ={base, fill=green!14},
  t2/.style    ={base, fill=yellow!32},
  t3/.style    ={base, fill=violet!16},
  t4/.style    ={base, fill=cyan!16},
  edge/.style  ={->, thick},
  eexe/.style  ={->, thick, black!45},
  pg/.style    ={->, thick, green!50!black},
  py/.style    ={->, thick, orange!85!black},
  pk/.style    ={->, thick, black!50},
  sw/.style    ={font=\scriptsize\itshape, text=black!60},
  lbl/.style   ={font=\footnotesize\bfseries},
  att/.style   ={circle, draw, fill=yellow!70!orange, inner sep=0pt,
                 minimum size=3.1mm, font=\scriptsize},
]

\begin{scope}[local bounding box=Z]
  \node[prompt] (z0) {prompt};
  \node[pln, below=of z0] (z1) {APM};
  \node[t1, below=of z1] (z2) {tool\_call};
  \node[t1, below=of z2] (z3) {tool\_output};
  \node[disp, below=of z3] (z6) {final output};
  \node[att] at (z6.north east) {\checkmark};
  \draw[edge] (z0)--(z1); \draw[edge] (z1)--(z2); \draw[edge] (z2)--(z3);
  \draw[edge] (z3)--(z6);
\end{scope}

\begin{scope}[local bounding box=A, shift={(3.0,0)}]
  \node[prompt] (a0) {prompt};
  \node[pln, below=of a0] (ap) {APM};
  \node[t1, below=of ap] (a1) {tool\_call};
  \node[t1, below=of a1] (a2) {tool\_output};
  \node[t2, below=of a2] (a3) {tool\_call};
  \node[t2, below=of a3] (a4) {tool\_output};
  \node[disp, below=of a4] (ac) {final output};
  \node[att] at (ac.north east) {\checkmark};
  \draw[edge] (a0)--(ap); \draw[edge] (ap)--(a1); \draw[edge] (a1)--(a2);
  \draw[edge] (a2)--(a3); \draw[edge] (a3)--(a4); \draw[edge] (a4)--(ac);
\end{scope}

\begin{scope}[local bounding box=B, shift={(12.2,0)}]
  \node[prompt] (b0) at (0,0) {prompt};
  \node[pln, below=of b0] (bp) {APM};
  \node[t1, below left=6mm and -2mm of bp]  (bl1) {tool\_call};
  \node[t2, below right=6mm and -2mm of bp] (br1) {tool\_call};
  \node[t1, below=3mm of bl1] (bl2) {tool\_output};
  \node[t2, below=3mm of br1] (br2) {tool\_output};
  \draw[edge] (b0)--(bp);
  \draw[edge] (bp)-- node[left=1pt,font=\scriptsize] {subagent-1} (bl1);
  \draw[edge] (bp)-- node[right=1pt,font=\scriptsize] {subagent-2} (br1);
  \draw[edge] (bl1)--(bl2);
  \draw[edge] (br1)--(br2);
  \node[disp] (bre) at ($($(bl2.south)!0.5!(br2.south)$)+(0,-1.4)$) {final output};
  \node[att] at ($(bre.north east)+(-0.35,0)$) {\checkmark};
  \draw[edge, draw=orange!70!black] (bl2.south) .. controls +(0,-6mm) and +(-6mm,6mm) .. (bre.north west);
  \draw[edge, draw=orange!70!black] (br2.south) .. controls +(0,-6mm) and +(6mm,6mm)  .. (bre.north east);
  \begin{scope}[on background layer]
    \node[fit=(bl1)(bl2), draw=black!30, dashed, rounded corners, inner sep=3pt] {};
    \node[fit=(br1)(br2), draw=black!30, dashed, rounded corners, inner sep=3pt] {};
  \end{scope}
\end{scope}

\begin{scope}[local bounding box=C, shift={(17.0,0)}]
  \node[prompt] (c0) at (0,0) {prompt};
  \node[pln, below=of c0] (cp) {APM};
  \node[t1, below left=6mm and -2mm of cp]  (cl1) {tool\_call};
  \node[t2, below right=6mm and -2mm of cp] (cr1) {tool\_call};
  \node[t1, below=3mm of cl1] (cl2) {tool\_output};
  \node[t3, below=3mm of cl2] (cl3) {tool\_call};
  \node[t3, below=3mm of cl3] (cl4) {tool\_output};
  \node[t2, below=3mm of cr1] (cr2) {tool\_output};
  \node[t4, below=3mm of cr2] (cr3) {tool\_call};
  \node[t4, below=3mm of cr3] (cr4) {tool\_output};
  \draw[edge] (c0)--(cp);
  \draw[edge] (cp)-- node[left=1pt,font=\scriptsize] {subagent-1} (cl1);
  \draw[edge] (cp)-- node[right=1pt,font=\scriptsize] {subagent-2} (cr1);
  \draw[edge] (cl1)--(cl2); \draw[edge] (cl2)--(cl3); \draw[edge] (cl3)--(cl4);
  \draw[edge] (cr1)--(cr2); \draw[edge] (cr2)--(cr3); \draw[edge] (cr3)--(cr4);
  \node[disp] (cre) at ($($(cl4.south)!0.5!(cr4.south)$)+(0,-1.4)$) {final output};
  \node[att] at ($(cre.north east)+(-0.35,0)$) {\checkmark};
  \draw[edge, draw=orange!70!black] (cl4.south) .. controls +(0,-6mm) and +(-6mm,6mm) .. (cre.north west);
  \draw[edge, draw=orange!70!black] (cr4.south) .. controls +(0,-6mm) and +(6mm,6mm)  .. (cre.north east);
  \begin{scope}[on background layer]
    \node[fit=(cl1)(cl4), draw=black!30, dashed, rounded corners, inner sep=3pt] {};
    \node[fit=(cr1)(cr4), draw=black!30, dashed, rounded corners, inner sep=3pt] {};
  \end{scope}
\end{scope}

\begin{scope}[local bounding box=D, shift={(7.2,0)}]
  \node[prompt] (d0) at (0,0) {prompt};
  \node[pln, below=of d0] (dp) {APM};
  \node[t1, below=of dp] (d1) {tool\_call};
  \node[t2, below=of d1] (d2) {tool\_call};
  \node[t1, below=of d2] (d4) {tool\_output};
  \node[t2, below=of d4] (d5) {tool\_output};
  \node[disp, below=of d5] (dc) {final output};
  \node[att] at ($(dc.north east)+(-0.15,0)$) {\checkmark};
  \draw[edge] (d0)--(dp);
  \draw[edge] (dp)--(d1);
  \draw[edge] (d1)--(d2);
  \draw[edge] (d2)--(d4);
  \draw[edge] (d4)--(d5);
  \draw[edge] (d5)--(dc);
  \draw[pg] (dp.west) to[bend left=45] (d0.west);
  \draw[py] (dp.east) to[bend right=45] (d0.east);
  \draw[pg] (d1.west) to[bend left=45] (dp.west);
  \draw[pg] (d4.west) to[bend left=50] (d1.west);
  \draw[pg] (dc.west) to[bend left=50] (d4.west);
  \draw[py] (d2.east) to[bend right=50] (dp.east);
  \draw[py] (d5.east) to[bend right=50] (d2.east);
  \draw[py] (dc.east) to[bend right=45] (d5.east);
  \node[rotate=90,  font=\scriptsize, text=green!50!black]  at (-1.5,-3.4) {Hash Chain for Task-1};
  \node[rotate=90, font=\scriptsize, text=orange!85!black] at (1.5,-3.4)  {Hash Chain for Task-2};
\end{scope}

\coordinate (legrow) at ($(current bounding box.south)+(0,-0.5)$);
\node[t1, anchor=west, minimum width=8mm, minimum height=4mm] (lg1)
      at ($(current bounding box.west |- legrow)+(0.05,0)$) {};
\node[anchor=west, font=\scriptsize] (lg1t) at (lg1.east) {Task-1};
\node[t2, anchor=west, minimum width=8mm, minimum height=4mm] (lg2) at ($(lg1t.east)+(0.45,0)$) {};
\node[anchor=west, font=\scriptsize] (lg2t) at (lg2.east) {Task-2};
\node[t3, anchor=west, minimum width=8mm, minimum height=4mm] (lg3) at ($(lg2t.east)+(0.45,0)$) {};
\node[anchor=west, font=\scriptsize] (lg3t) at (lg3.east) {Task-3};
\node[t4, anchor=west, minimum width=8mm, minimum height=4mm] (lg4) at ($(lg3t.east)+(0.45,0)$) {};
\node[anchor=west, font=\scriptsize] (lg4t) at (lg4.east) {Task-4};

\coordinate (figbot) at (current bounding box.south);
\node[lbl, anchor=north, align=center] at ($(Z.south |- figbot)+(0,-0.15)$)
      {(a) Single-task\\single-agent};
\node[lbl, anchor=north, align=center] at ($(A.south |- figbot)+(0,-0.15)$)
      {(b) Multi-task\\single-agent};
\node[lbl, anchor=north, align=center] at ($(D.south |- figbot)+(0,-0.15)$)
      {(c) Multi-task\\single-agent (interleaved)};
\node[lbl, anchor=north, align=center] at ($(B.south |- figbot)+(0,-0.15)$)
      {(d) Multi-task multi-agent\\(\#tasks $=$ \#agents)};
\node[lbl, anchor=north, align=center] at ($(C.south |- figbot)+(0,-0.15)$)
      {(e) Multi-task multi-agent\\(\#tasks $>$ \#agents)};

\end{tikzpicture}%
}
\caption{\textbf{Overview of Agentic Workloads:} AEGs across task and agent settings where node colors distinguish tasks: (a) single-task, single-agent (ST-SA); (b) multi-task, single-agent (sequential; MT-SA); (c) multi-task, single-agent (interleaved), with curved arcs showing per-task hash chains and skip-arcs showing interleaving. (d)--(e) Multi-task, multi-agent where APM bifurcates into one branch per sub-agent for (d) each task, or (e) multiple tasks.}
\label{fig:scalability}
\end{figure*}
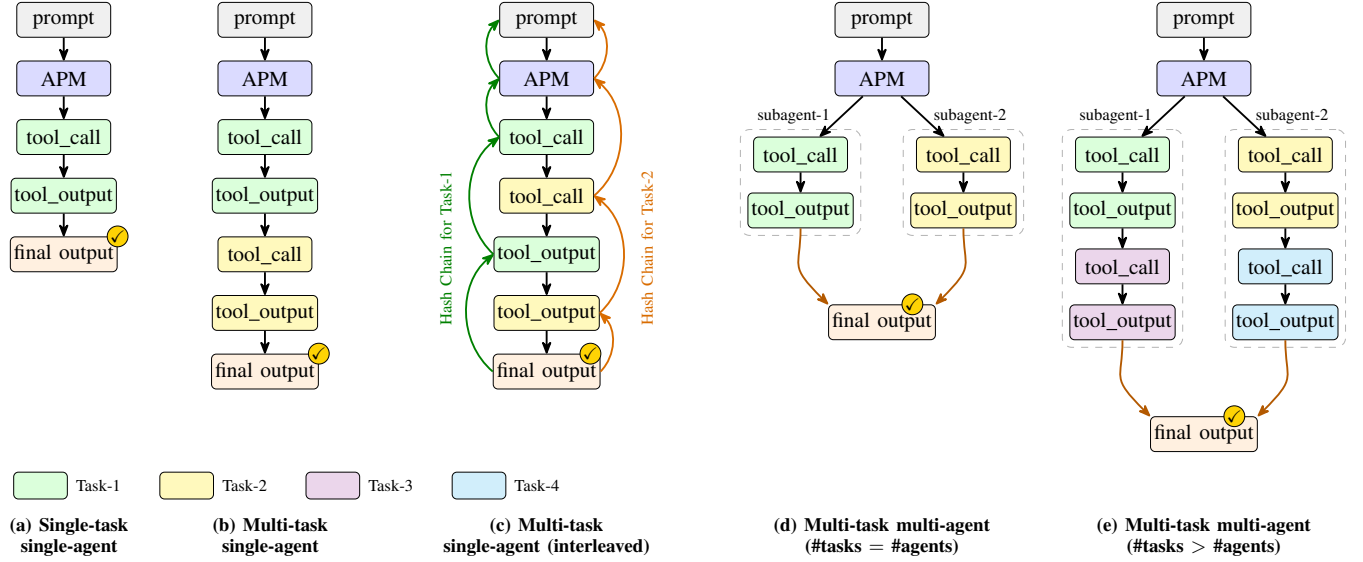

\noindent\textbf{\underline{Agentic Workloads:}} A run is parameterized by the number of tasks $t$ and executor sub-agents $m$ ($1 \le m \le t$). 
Varying $(t,m)$ changes the shape of the AEG, from a single chain to a directed graph. We consider five cases (Figure~\ref{fig:scalability}):
\begin{enumerate}[leftmargin=*,topsep=2pt,itemsep=1pt]
\item \textbf{Single-task, single-agent (ST-SA; $t=1,\ m=1$)}: One executor handles one task, and the AEG is a single permitted path. This is the simplest baseline case.
\item \textbf{Multi-task, single-agent sequential (MT-SA seq; $t>1,\ m=1$)}: One executor sequentially handles $t$ tasks, where the trajectories are concatenated, and the executor finishes each task's tool calls before handling the next task. The AEG is one chain whose trajectory is the ordered concatenation of the $t$ per-task trajectories.
\item \textbf{Multi-task, single-agent interleaved (MT-SA int; $t>1,\ m=1$)}: One executor handles $t$ tasks whose tool calls are \emph{interleaved} rather than run to completion in blocks. The AEG is a single chain, but the permitted trajectory interleaves tool calls from different tasks.
\item \textbf{Multi-task, multi-agent (MT-MA, $t>m$)}: The planner distributes $t$ tasks among $m$ sub-agents. Here, the AEG splits into $m$ branches of uneven length.
\item \textbf{Multi-task, multi-agent (MT-MA, $t=m > 1$)}: The planner spawns one sub-agent per task and the AEG splits into $t$ branches, one per task.
\end{enumerate}

\begin{figure}[!t]
\centering
\resizebox{0.82\columnwidth}{!}{%
\begin{tikzpicture}[
  >={Stealth[round]},
  font=\small,
  node distance=3.4mm and 8mm,
  base/.style={draw, rounded corners=2pt, minimum height=5.2mm, minimum width=13mm,
               align=center, inner sep=1.5pt},
  prompt/.style={base, fill=black!6},
  plan/.style  ={base, fill=blue!14},
  call/.style  ={base, fill=green!8},
  outn/.style  ={base, fill=green!14},
  blockn/.style={base, fill=red!10, draw=red!60},
  ghost/.style ={base, dashed, fill=none, text=black!45, draw=black!35},
  disp/.style  ={base, fill=orange!12},
  dev/.style   ={base, densely dotted, thick, draw=red!70, fill=orange!8},
  ext/.style   ={base, dashed, draw=red!60, fill=gray!12},
  pol/.style   ={base, fill=gray!12},
  chk/.style   ={base, fill=yellow!22, draw=orange!75!black},
  att/.style   ={circle, draw, fill=yellow!70!orange, inner sep=0pt,
                 minimum size=3.1mm, font=\scriptsize},
  edge/.style  ={->, thick},
  gedge/.style ={->, dashed, draw=black!40},
  lbl/.style   ={font=\footnotesize\bfseries},
]


\begin{scope}[local bounding box=B, shift={(4.6,-6.7)}]
  \node[prompt] (b0) at (0,0) {prompt};
  \node[plan, below=of b0] (b1) {APM};
  \node[call, below=of b1] (b2) {tool\_call};
  \node[outn,  below=of b2] (b3) {tool\_output};
  \node[blockn, below=of b3] (bb) {interrupt};
  \node[ghost, below=of bb] (b4) {tool\_call};
  \node[ghost, below=of b4] (b5) {(pending)};
  \node[att] at (bb.north east) {\checkmark};
  \draw[edge] (b0)--(b1); \draw[edge] (b1)--(b2); \draw[edge] (b2)--(b3);
  \draw[edge, draw=red!70] (b3)--(bb);
  \draw[gedge] (bb)--(b4); \draw[gedge] (b4)--(b5);
  \draw[red, line width=0.9pt, decorate,
        decoration={zigzag, segment length=2mm, amplitude=0.6mm}]
        ($(b3.south west)+(0.1,-0.2)$) -- ($(b3.south east)+(-0.1,-0.2)$);
\end{scope}

\begin{scope}[local bounding box=D, shift={(4.6,0)}]
  \node[prompt] (d0) at (0,0) {prompt};
  \node[plan, below=of d0] (d1) {APM};
  \node[call, below=of d1] (d2) {tool\_call};
  \node[outn, below=of d2] (d3) {tool\_output};
  \node[ext] (dx) at ($(d3.west)+(-1.35,0.65)$) {untrusted\\source};
  \node[ghost] (dp) at ($(d3.south)+(-0.95,-0.8)$) {tool\_call\\(planned)};
  \node[dev]   (dv) at ($(d3.south)+(0.95,-0.8)$)  {tool\_call\\(deviated)};
  \node[chk, below=3mm of dv] (di) {$\Gamma_{\mathrm{traj}}$};
  \node[blockn, below=3mm of di] (db) {blocked};
  \node[att] at (db.north east) {\checkmark};
  \draw[edge] (d0)--(d1); \draw[edge] (d1)--(d2); \draw[edge] (d2)--(d3);
  \draw[->, dashed, black!60] (d2.west) to[bend right=15]
        node[above left=0pt and -2pt, font=\scriptsize, text=black!60] {fetch} (dx.north);
  \draw[edge, draw=red!70] (dx.south east) --
        node[midway, sloped, below=1pt, font=\scriptsize, text=red!70] {injected} (d3.west);
  \draw[gedge] (d3)--(dp);
  \draw[edge, draw=red!70] (d3)--(dv);
  \draw[edge, draw=red!70] (dv)--(di);
  \draw[edge, draw=red!70] (di)--(db);
\end{scope}

\begin{scope}[local bounding box=E, shift={(0,-1.1)}]
  \node[prompt] (e0) at (0,0) {prompt};
  \node[plan, below=of e0] (e1) {APM};
  \node[call, below=of e1] (e2) {tool\_call};
  \node[chk, below=of e2] (rc) {$\Gamma_{\mathrm{reg}}$};
  \node[blockn, below=of rc] (eb) {blocked};
  \node[pol] (pd) at ($(rc.east)+(1.15,0)$) {regulation\\documents};
  \node[att] at (eb.north east) {\checkmark};
  \draw[edge] (e0)--(e1); \draw[edge] (e1)--(e2);
  \draw[edge] (e2)--(rc);
  \draw[->, dashed, black!60] (pd.west) -- (rc.east);
  \draw[edge, draw=red!70] (rc)-- node[left=1pt, font=\scriptsize, text=red!70] {Fail} (eb);
\end{scope}

\begin{scope}[local bounding box=F, shift={(0,-7.5)}]
  \node[prompt] (f0) at (0,0) {prompt};
  \node[plan, below=of f0] (f1) {APM};
  \node[call, below=of f1] (f2) {tool\_call\\(system call)};
  \node[chk, below=of f2] (fs) {$\Gamma_{\mathrm{sys}}$\\(seccomp)};
  \node[blockn, below=of fs] (fb) {blocked};
  \node[att] at (fb.north east) {\checkmark};
  \draw[edge] (f0)--(f1); \draw[edge] (f1)--(f2);
  \draw[edge] (f2)--(fs);
  \draw[edge, draw=red!70] (fs)--(fb);
\end{scope}

\coordinate (r1y) at ($(D.south)+(0,-0.2)$);
\node[lbl, anchor=north] at (e0.south |- r1y) {(a) Blocked: $\Gamma_{\mathrm{reg}}$};
\node[lbl, anchor=north] at (d0.south |- r1y) {(b) Blocked: $\Gamma_{\mathrm{traj}}$};
\coordinate (r2y) at ($(B.south)+(0,-0.2)$);
\node[lbl, anchor=north] at (f0.south |- r2y) {(c) Blocked: $\Gamma_{\mathrm{sys}}$};
\node[lbl, anchor=north] at (b0.south |- r2y) {(d) Interrupt: $\Gamma_{\mathrm{int}}$};

\coordinate (legrow) at ($(current bounding box.south)+(0,-0.28)$);
\node[att, anchor=west] (lg) at ($(current bounding box.west |- legrow)+(0.3,0)$) {\checkmark};
\node[anchor=west, font=\scriptsize] (lgt) at (lg.east) {Chain Attestation};
\node[plan, anchor=west, minimum width=8mm, minimum height=4mm] (lg3) at ($(lgt.east)+(0.3,0)$) {};
\node[anchor=west, font=\scriptsize] (lg3t) at (lg3.east) {By planner};
\node[call, anchor=west, minimum width=8mm, minimum height=4mm] (lg4) at ($(lg3t.east)+(0.3,0)$) {};
\node[anchor=west, font=\scriptsize] (lg4t) at (lg4.east) {By executor};
\coordinate (legrow2) at ($(legrow)+(0,-0.42)$);
\node[ghost, anchor=west, minimum width=8mm, minimum height=4mm] (lg2) at (lg.west |- legrow2) {};
\node[anchor=west, font=\scriptsize] (lg2t) at (lg2.east) {In-memory only, never persisted};
\draw[red, line width=0.9pt, decorate,
      decoration={zigzag, segment length=2mm, amplitude=0.6mm}]
      ($(lg2t.east)+(0.3,0)$) -- ($(lg2t.east)+(1.0,0)$);
\node[anchor=west, font=\scriptsize] at ($(lg2t.east)+(1.1,0)$) {Interrupt};

\end{tikzpicture}%
}
\caption{\textbf{Types of Nodes in AEG:} Blue node is the planner-generated APM; green nodes record the executor’s actions. Each tool call passes through \RefMon, which admits only compliant calls and records an attested \texttt{check} node. Each node embeds its parent’s hash, with one chain-head attestation covering the AEG. (a)--(c) Violations of $\Gamma_{\mathrm{reg}}$, $\Gamma_{\mathrm{traj}}$, or $\Gamma_{\mathrm{sys}}$ produce an attested \texttt{blocked} node without \texttt{tool\_output}; (d) an interrupt produces an attested \texttt{interrupt} node.}
\label{fig:tree}
\end{figure}
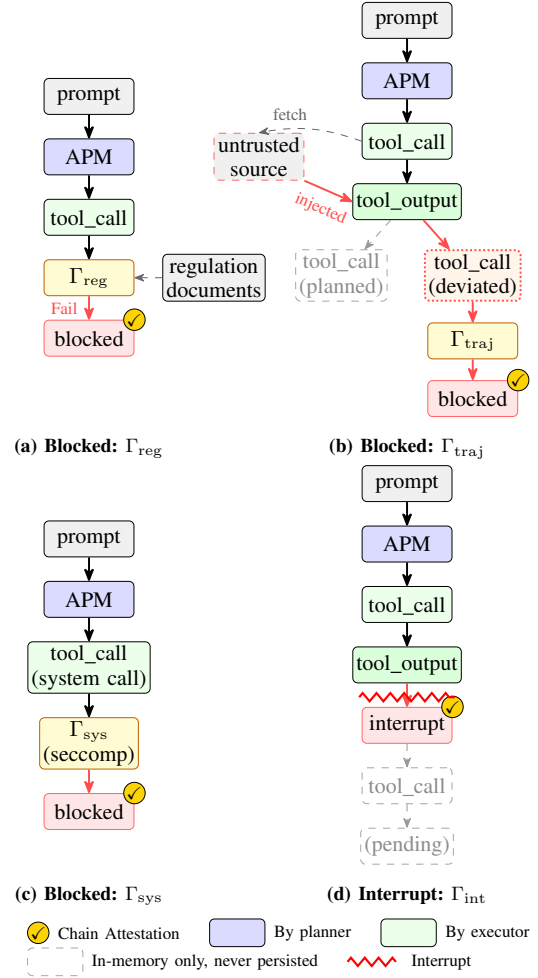

\noindent\textbf{\underline{AEG Topologies across Workloads:}} Each node in AEG includes its type, its parent, the acting executor, a content payload, an \texttt{allowed} flag with a \texttt{block\_reason}, a per-node \emph{measurement}, and a list of attestations. 
Every executor action creates a \texttt{tool\_call} node, storing the position in the trajectory, the expected call and its hash, and the actual call and its hash, so the decision is self-describing and re-computable for the verifier.
The control flow then goes through the checks, done by \RefMon, where each check is recorded as an attested \texttt{check} child. The check has the following outcomes: (i) if it passes, the tool executes and its result is appended as a \texttt{tool\_output} child in the AEG; (ii) if it fails, a \texttt{blocked} node is appended to the failing \texttt{check} and the tool is not executed; (iii) if interrupted, an \texttt{interrupt} node is appended to the AEG. The various node types and various types of checks are shown in Figure~\ref{fig:tree}.
We discuss what the AEG looks like for different workloads:
\begin{itemize}[leftmargin=*,itemsep=0pt,topsep=0pt]
\item \textbf{Single-agent Workloads:} For ST-SA and MT-SA, one executor handles all the computation and the AEG is a \emph{linear} chain of attested trajectory (Figure~\ref{fig:scalability}a,b). Interleaving (MT-SA int) does not change this shape as the task-switching across tools only permutes the order of the nodes, not the number of paths (Figure~\ref{fig:scalability}c). 

\item \textbf{Multi-agent Workloads:} The planner decomposes the task into sub-tasks, recorded as an attested \texttt{agent\_step} (\Cref{att:disp}), and dispatches one sub-agent per sub-task, bifurcating the AEG.
Each \texttt{agent\_branch} is an independent subgraph with its own APM trajectory and \texttt{tool\_call} chain; branches run in parallel without sharing nodes (Figure~\ref{fig:scalability}d,e). 
They recombine at the root, where the shared parent binds the complete branch set through the structure root. Thus, aggregation is a graph-level operation, and the complete AEG is serialized and verified as a single attested object.
\end{itemize}

\noindent\textbf{\underline{Optimizing Attestations:}} \method uses two optimizations that lower the attestation cost:
\begin{itemize}[leftmargin=*,itemsep=0pt,topsep=0pt]
\item \textbf{One Attestation per AEG:} Rather than one hardware attestation per node, a single attestation covers the AEG summary $(c_N,s)$, which already commits to every node's measurement and to the topology. 
The agent cannot tamper with the AEG, since the executor never touches the evidence directly, and every action passes through \RefMon. 
An executor therefore cannot add, drop, or alter a node without \RefMon recording and blocking it. Any change to AEG breaks the hash comparisons (\S\ref{sec:verification}).
\item \textbf{In-TEE Signing Key:} Instead of invoking the hardware for every attestation, \method generates a signing key inside the TD at startup and binds its public key into the hardware attestation. Each AEG is then signed with this key and costs one in-memory signature, while the single hardware attestation certifies the key that produced every signature.
\end{itemize}

\subsection{Verification Process}\label{sec:verification}

Verification checks the AEG against the APM and confirms that (a) the APM was derived from the prompt, and (b) every action was checked against it. 
A single hardware attestation covers the entire AEG, allowing any verifier to check the serialized artifact offline using the artifact and one quote.
A verifier accepts the AEG if all the following pass:
\begin{enumerate}[leftmargin=*]
\item \textbf{Evidence:} Verify the Intel TDX quote and NVIDIA GPU token, and confirm that the quote's \texttt{report\_data} matches the recomputed AEG summary containing the chain head $c_N$ and structure root $s$. The chain binds all node contents, while $s$ binds the AEG topology.

\item \textbf{Chain:} Recompute each node's measurement $m_i$ and fold them in commit order. The resulting head must equal the value attested by the hardware, detecting any missing, reordered, or modified node.

\item \textbf{Binding Inference:} For each \texttt{proof\_of\_inference} (Appendix~\ref{app:attestations}: \Cref{att:inf}), verify that \texttt{weights\_hash} matches the claimed model, including the planner, executor, and the judge used for checking regulations.

\item \textbf{Policy Consistency:} Recompute tool call hashes and guards from the APM. Each admitted \texttt{tool\_call} must match the expected tool call at its position in $\traj$, and each failed guard must have a corresponding \texttt{blocked} node.

\item \textbf{Topology:} Recompute $s$ and verify that the AEG has the expected topology: each \texttt{check} is attached to its corresponding \texttt{tool\_call}, each \texttt{agent\_branch} is attached to the root, and every sub-task dispatched by the planner has exactly one branch. This detects any missing or inserted node or sub-agent branch.
\end{enumerate}

\noindent\textbf{\underline{Verifying AEG Types:}} The above verification distinguishes three cases: (a) an AEG whose recomputed head and structure root equal the values attested by the hardware is complete and untampered; (b) an internally consistent AEG with no valid hardware attestation is an unfinished run, admitted only as an integrity-checked prefix; and (c) any measurement mismatch, sequence gap, broken predecessor link, or invalid attestation indicates tampering. 


\noindent\textbf{\underline{Blocking is Sufficient:}}
Prior prompt-injection defenses aim to preserve task utility despite malicious prompts, and simply blocking deviation degrades task utility~\cite{kolluri2026optimizing}. In our setting, however, the fraction of blocked actions provides sufficient evidence for verifiers to guide deployment decisions. For example, verifiers could require the blocked-action rate across AEGs and prompts to remain below a threshold for deployment, while recommending stronger defenses or model fine-tuning to improve compliance.
\section{Experimental Setup}\label{sec:setup}

\noindent\textbf{\underline{Toolkit:}} We build a framework to simulate diverse agentic workloads, including multi-agent spawning and tool calls implemented as functions. We use InjecAgent~\cite{zhan2024injecagent}, a benchmark for prompt injection attacks that specifies $40$ toolkits (e.g., \texttt{Gmail}, \texttt{Amazon}, \texttt{GitHub}, \texttt{Slack}, \texttt{Teladoc}, \texttt{Terminal}), exposing $333$ tools in total, with between $1$ and $13$ tools per toolkit ($8.3$ on average). However, InjecAgent tools are represented as JSON inputs rather than executable tool calls. 
We convert these JSON specifications into functions that return placeholder responses, enabling realistic tool-calling behavior while preserving the benchmark's scenarios.
For tool-call deviation, the executor is given malicious instructions (via prompt injection) to force a deviation.

\noindent\textbf{\underline{Datasets and Splits:}}
We generate a common pool of flows across toolkits, and split it into ``block'' and ``allow''.
\begin{itemize}[leftmargin=*,topsep=0pt,itemsep=0pt]
\item \textbf{Common Flow Pool:} All experiments use a pool of $400$ flows, comprising $K{=}10$ flows per toolkit across the $40$ toolkits. Each flow is annotated for one of four independent guards enforced by \RefMon: (i) \emph{regulation}, where the flow is realized by two actions, a \emph{read} that retrieves some data and a \emph{disclosure} that sends it to an outside recipient (e.g., by e-mail), and the guard is applied at the disclosure; (ii) \emph{unauthorized tool call}, where a malicious instruction in previous output forces a deviation; (iii) \emph{unauthorized system call}, where the executor runs \texttt{Terminal.Execute} that the APM does not permit; and (iv) \emph{interruption}, where the flow must be stopped.

\item \textbf{Generation:} We modify the above pool to include disclosures and unauthorized tool/system calls. For regulation, the compliant and non-compliant flows are identical except for whether the recipient is permitted and consent was given. As examples, we use GDPR and HIPAA on the $15$ toolkits that can send data to an outside recipient. For unauthorized tool calls, we introduces deviations using prompt injections. We follow InjecAgent where the executor reads a tool output, which differs only in the prompt injected instruction.

\item \textbf{Splits:} The data is split into $75$ positive and $75$ negative flows for regulation, and $200$ of each for prompt injection, unauthorized system calls, and interrupts.
Table~\ref{tab:workloads} reports how each workload partitions this pool into bundles. Bundling happens at runtime and adds no new data, so every workload evaluates the same flows.
\end{itemize}
We count the dataset in \emph{flows} (Table~\ref{tab:workloads}): ST-SA is one flow per bundle, MT-SA puts $t$ flows in one branch, and MT-MA spreads them over $m$ branches, so every workload re-partitions the same flows and the totals are comparable.

\begin{table}[htb]
\centering
\small
\setlength{\tabcolsep}{4pt}
\caption{\textbf{Dataset Overview:} Each workload ($t$ tasks, $m$ sub-agents) partitions the information flow pool into bundles (``Bund.'') of $t$ flows.}
\label{tab:workloads}
\begin{tabular}{@{}llr rr rr@{}}
\bottomrule

\toprule
 & & & \multicolumn{2}{c}{Regulation} & \multicolumn{2}{c}{Tools / Sys. / Int} \\
\cmidrule(lr){4-5}\cmidrule(lr){6-7}
 & $t$ & $m$ & Bund. & Flows & Bund. & Flows \\
\bottomrule

\toprule
ST-SA     & $1$  & $1$  & $150$ & $150$ & $400$ & $400$ \\
\midrule
MT-SA-seq  & $2$  & $1$  & $75$ & $150$ & $200$ & $400$ \\
           & $4$  & $1$  & $37$ & $148$ & $100$ & $400$ \\
           & $6$  & $1$  & $25$ & $150$ & $\phantom{0}66$ & $396$ \\
\midrule
MT-SA-int  & $2$  & $1$  & $75$ & $150$ & $200$ & $400$ \\
           & $4$  & $1$  & $37$ & $148$ & $100$ & $400$ \\
           & $6$  & $1$  & $25$ & $150$ & $\phantom{0}66$ & $396$ \\
\midrule
MT-MA      & $2$  & $2$  & $75$ & $150$ & $200$ & $400$ \\
           & $4$  & $2$  & $37$ & $148$ & $100$ & $400$ \\
           & $4$  & $4$  & $37$ & $148$ & $100$ & $400$ \\
           & $6$  & $2$  & $25$ & $150$ & $\phantom{0}66$ & $396$ \\
           & $6$  & $4$  & $25$ & $150$ & $\phantom{0}66$ & $396$ \\
           & $6$  & $6$  & $25$ & $150$ & $\phantom{0}66$ & $396$ \\
\bottomrule

\toprule
\end{tabular}
\end{table}

\noindent\textbf{\underline{Models:}} We use different models for the planner and executors. The planner uses \modelverb{Qwen3.5-9B} with thinking mode enabled for chain-of-thought reasoning. To avoid model-family bias, executors use models from three families: \modelverb{Llama-3.1-8B}, \modelverb{Gemma-3-12B}, and \modelverb{Ministral-3-14B}. We omit the ``-Instruct'' suffix for brevity, and all models are open-weight and publicly available. For compliance judges, we use a separate instance of \modelverb{Qwen3.5-9B} with reasoning enabled to improve accuracy.

\noindent\textbf{\underline{Evaluation Metrics:}} We describe the metrics for \RefMon's compliance effectiveness and for its efficiency.
\begin{itemize}[leftmargin=*]
\item \noindent\textbf{Effectiveness:} Treating a blocked flow as the positive case, we report the \emph{incorrectly blocked} rate (the false-positive rate) and the \emph{incorrectly allowed} rate.
For $\Gamma_{\mathrm{reg}}$, we score a flow only when the executor actually sends the protected data.
For $\Gamma_{\mathrm{traj}}$ and $\Gamma_{\mathrm{sys}}$, we score a flow only when the executor attempts an off-trajectory action.

\item\textbf{Efficiency:} We measure the following overheads:
\begin{itemize}
\item \emph{AEG:} building the attested execution graph and running the guard checks (hash comparison, action-tuple derivation, and node and hash-chain construction, with no model and no cryptography);
\item \emph{Meas:} hashing the attested payloads;
\item \emph{HW:} the time the Intel TDX host and the GPU spend producing the quote and the GPU evidence; and
\item \emph{Sign:} the in-TEE signature over each execution record.
\end{itemize}
Overheads are a percentage of the baseline and reported as the mean and standard deviation over five runs. 
The one-time hash of the model weights is an offline cost paid once at the beginning. 
For multi-agent workloads, sub-agents run in parallel, so a bundle's baseline is the inference time of its slowest branch.
\end{itemize}
\noindent\textbf{Note on TEE Overhead:} The above costs assume workloads are already executed \emph{inside} the TEE, and we report the cost added by \method.
This is along the lines of prior work~\cite{duddu2024laminator,chantasantitam2026pal}.
The overhead of running models in TEE, is not introduced by \method.
Prior benchmarks report that these overheads are negligible, less than 7\%~\cite{chrapek2025confidential,zhu2024confidential,wang2026benchmarking}.
\section{Evaluation}\label{sec:evaluation}

We present the compliance effectiveness across various guards (\S\ref{sec:evalEffectiveness}), attestation overhead (\S\ref{sec:evalEfficiency}), and a security analysis for \method (\S\ref{sec:security}).

\subsection{Compliance Effectiveness}\label{sec:evalEffectiveness}

We discuss the effectiveness of \RefMon in checking for deviations from regulations, expected tools and system calls\footnote{\textbf{Note:} Our goal is not to improve agent compliance, but to assess whether compliance mechanisms correctly detect deviations to inform regulators.}:
\begin{itemize}[leftmargin=*,itemsep=0pt,topsep=0pt]
\item \textbf{Regulation Guard ($\Gamma_{\mathrm{reg}}$):} 
\RefMon uses a judge to check whether the executor's trajectory is compliant under regulations (e.g., GDPR and HIPAA).  
We evaluate the three executors against the full regulation pool over all thirteen workloads, scoring only flows in which the executor actually discloses protected information. 
For \modelverb{Llama-3.1-8B}, the monitor blocks $100\%$ of the non-compliant flows and allows $96\%$ of the compliant ones. 
For \modelverb{Gemma-3-12B}, the monitor blocks $81.8\%$ of the non-compliant flows and allows $92.5\%$ of the compliant ones. 
For \modelverb{Ministral-3-14B}, the monitor allows $90\%$ of the compliant flows and blocks $100\%$ of the non-compliant ones.
The numbers are stable across workloads which change the execution order but not the evaluated flows. 

\item\textbf{Trajectory Guard ($\Gamma_{\mathrm{traj}}$):} Any deviations in tool calls due to prompt injections should be caught by \RefMon by matching them against the APM. 
We evaluate the three executors, scoring each attempted deviation against the expected trajectory in the APM. 
As conjectured, correctly-blocked and correctly-allowed rates are $100\%$ for every model and workload.

\item \textbf{System Call Guard ($\Gamma_{\mathrm{sys}}$):} Once a tool call attempts an action outside the APM, \RefMon must block it. 
To validate this we evaluate the three executors, and we find that the correctly-blocked and correctly-allowed rates are $100\%$.
However, a compromised or exploited tool can issue a system call from \emph{inside} its own execution (\emph{unmediated}), which \RefMon cannot inspect. 
These are stopped only by the OS sandbox (seccomp-BPF + Landlock) provisioned per the APM's least-privilege policy.
Measuring this over the $100$-case corpus confirms the defense-in-depth design: the monitor alone prevents $74\%$ of deviations (all $100\%$ mediated, but only the $48\%$ baseline of unmediated), while adding the OS sandbox reaches $100\%$ prevention (mediated and
unmediated). Thus, $\Gamma_{\mathrm{sys}}$ requires both the monitor for mediated violations and the sandbox for unmediated ones.

\item \textbf{Stop Guard ($\Gamma_{\mathrm{int}}$):} Interruption is a control decision \RefMon makes directly from the action (a timeout, a resource cap, or an external stop signal). Therefore, \RefMon should enforce interruption with perfect soundness and completeness.
We empirically confirm $100\%$ correctly-blocked and correctly-allowed rates, indicating that a deterministic \RefMon decision is sound and complete.
\end{itemize}

\begin{takeaway}
\textbf{Takeaway:} \method shows effective compliance as \RefMon prevents any deviation across all models and workloads, and regulators can use the fraction of blocked responses to determine whether the agent can be deployed.
\end{takeaway}

\subsection{Efficiency}\label{sec:evalEfficiency}

We present the overhead of the various computations in APM generation and for the individual guards.

\begin{figure}[htb]
\centering
\includegraphics[width=\columnwidth]{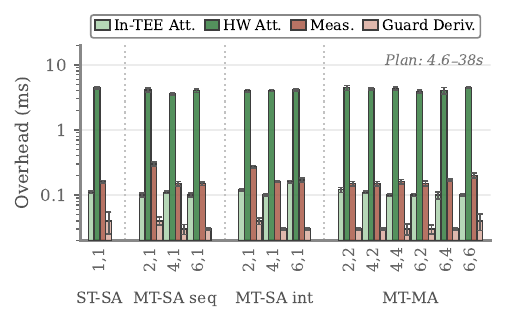}
\caption{\textbf{APM Generation Overhead:} We report the overhead of measurement, guard derivation, and hardware and in-TEE attestations across different workloads indicated as (\#task, \#agents), as a percentage of the baseline computation for planning. \emph{Meas.} is measurement, \emph{Att.} is attestation, and \emph{Deriv.} is derivation. $y$-axis is log-scaled.}
\label{fig:apmrelative}
\end{figure}

\noindent\textbf{\underline{APM Generation:}} The cost for generating the APM should be close to generating a standard plan given a prompt by the planner.
The trajectory $\traj$ is already produced by the planner; the only additions are deriving the guards $\Gamma$, assembling and hashing the APM, and binding it in a single attestation.
We report the results in Figure~\ref{fig:apmrelative} covering the overhead of deriving the APM's guards from the plan (``Guard Derivation''), APM serialization and measurement (``Meas.''), the hardware attestation over the manifest measurement (``HW Attestation''), and the in-TEE attestation, each as a percentage of the planning cost.
We find that all the operations for APM are negligible relative to planner inference.

Measurement is between 0.15–0.30ms, while the HW attestation takes between 3.5–4.5ms. 
In-TEE attestation takes around 0.10–0.16ms, which is significantly lower than the hardware attestation. The guard derivation takes around 0.03–0.04ms and the total overhead for the APM is around 4.5 ms, which is $\leq 0.1\%$ of the baseline cost for planning.
The cost of generating and attesting the APM is essentially invariant to the workload.
Because the planning baseline grows with workload complexity (from $4.6$s for \textsc{ST-SA} to $37.7$s for the largest \textsc{MT-MA}), the \emph{relative} overhead falls by roughly $8\times$, from $\sim\!0.10\%$ to $\sim\!0.013\%$. 
MT-SA (seq.) and MT-SA (int.) give identical APM cost at matched $t$ (e.g. HW 4.08 vs 4.00 ms at (2,1)) which is expected, since interleaving changes only the order of execution and not the manifest.
The baseline cost for generating a plan increases from 4.6s (ST-SA) to $\sim$33s at $t{=}6$ single-agent and up to 37.7s at MT-MA (6,6), and MT-MA is cheaper per workload where parallelism helps: (6,2) $\approx$17s vs single-agent (6,1) $\approx$34s.
\begin{takeaway}
\textbf{Takeaway:} Per-prompt APM overhead is a fixed, negligible cost that becomes cheaper as workloads scale.
\end{takeaway}



\noindent\textbf{\underline{Overheads for Guards:}} HW Attestation ($\approx 4$ ms) and In-TEE Attestation ($\approx 0.1$ ms) remain constant across workloads, models, and guards, reflecting an $O(1)$ attestation cost per AEG. AEG cost varies with graph volume, dips for all $t=6$ workloads, and increases with model size (\modelverb{Ministral-3-14B} $>$ \modelverb{Llama-3.1-8B} $\gtrsim$  \modelverb{Gemma-3-12B}). Since the baseline cost also increases with model size, relative overhead is smallest for the largest model. Overall, overhead is below $1\%$ of baseline in every case.
\begin{itemize}[leftmargin=*,topsep=0pt,itemsep=0pt]
\item\textbf{Regulation Guard ($\Gamma_{\mathrm{reg}}$):} This is the costliest guard, as it invokes an LLM judge on every disclosing flow, which dominates node count. 
As shown in Figure~\ref{fig:regulation}, the hardware attestation is $3.7-4.6$ms and the signature $\approx0.1$ms as usual, while \emph{AEG} is $21-40$ms and measurement $4.9-9.1$ms (roughly twice the other guards) for a total of $30-53$ms against a large baseline of $29-231$s. 
Because that baseline is so large, its \emph{relative} overhead is in fact the lowest: $0.017-0.025\%$ in every single-agent workload and at most $0.125\%$ in the most parallel multi-agent cell.

\begin{figure}[htb]
\centering
\includegraphics[width=\columnwidth]{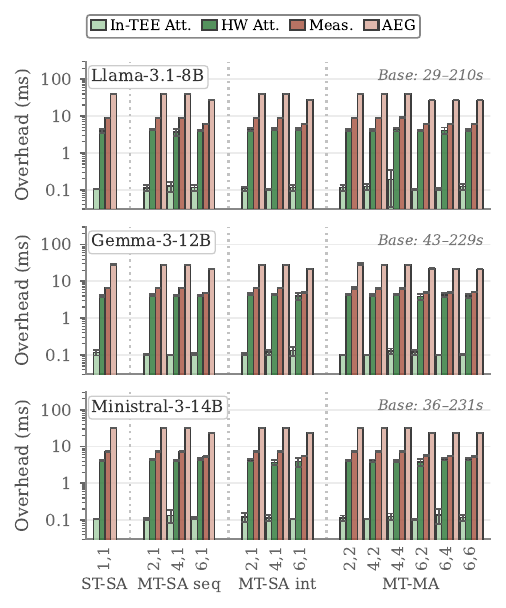}
\caption{\textbf{Regulation Guard:} We report AEG generation, measurement, and attestation overheads across workloads and models (both in-TEE and hardware attestations) as a percentage of the baseline computation cost. We indicate the workloads as (\#task, \#agents). $y$-axis is log-scaled.}
\label{fig:regulation}
\end{figure}

\item\textbf{Trajectory Guard ($\Gamma_{\mathrm{traj}}$):} Here, the guard is a fast hash comparison against the APM, with no LLM in the loop. We report the results in Figure~\ref{fig:trajectory}.
The hardware attestation is $3.2-4.7$ms and the signature $\approx0.1$ms, with \emph{AEG} $13-31$ms and measurement $3.0-6.3$ms, for a total of $19-41$ms over a baseline of $5.9-117$s. 
The full guard therefore costs $0.032-0.100\%$ of the baseline in every single-agent workload and at most $0.384\%$ in the most parallel multi-agent cell.

\begin{figure}[htb]
\centering
\includegraphics[width=\columnwidth]{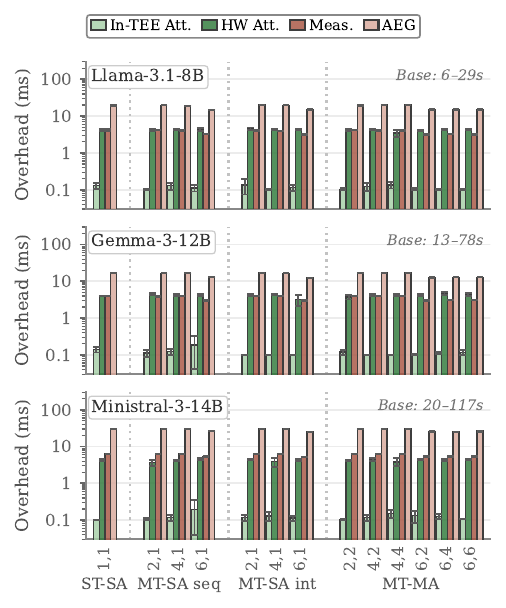}
\caption{\textbf{Trajectory Guard:} We report AEG generation, measurement, and attestation overheads across workloads and models (both in-TEE and hardware attestations) as a percentage of the baseline computation cost. We indicate the workloads as  (\#task, \#agents). $y$-axis is log-scaled.}
\label{fig:trajectory}
\end{figure}

\item\textbf{System Call Guard ($\Gamma_{\mathrm{sys}}$):} This is deterministic OS-level mediation, with cost close to $\Gamma_{\mathrm{traj}}$. We report the results in Figure~\ref{fig:oscheck}.
The hardware attestation is $3.3-4.7$ms and the signature $\approx0.1$ms, with \emph{AEG} $12-23$ms and measurement $2.8-4.8$ms, for a total of $19-32$ms over a baseline of $4.5-79$s, i.e., $0.029-0.114\%$ of the baseline for single-agent workloads and at most $0.521\%$ in the most parallel multi-agent cell.

\begin{figure}[htb]
\centering
\includegraphics[width=\columnwidth]{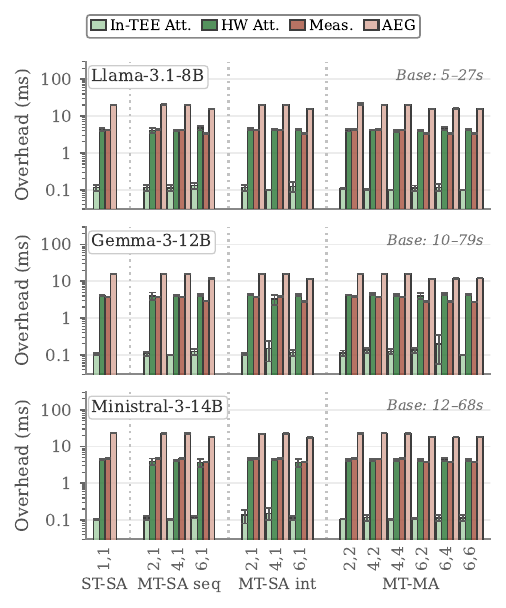}
\caption{\textbf{System Call Guard:} We report AEG generation, measurement, and attestation overheads across workloads and models (both in-TEE and hardware attestations) as a percentage of the baseline computation cost. We indicate the workloads as (\#task, \#agents). $y$-axis is log-scaled.}
\label{fig:oscheck}
\end{figure}

\item \textbf{Stop Guard ($\Gamma_{\mathrm{int}}$):} This is the cheapest guard, costing only graph construction plus the fixed attestation tax. We report the results in Figure~\ref{fig:interrupt}.
The hardware attestation is $3.4-4.6$ms and the signature $\approx0.1$ms, with \emph{AEG} $11-19$ms and measurement $2.6-3.9$ms, for a total of $18-27$ms over a baseline of $2.8-77$s. 
The guard costs $0.030-0.120\%$ of the baseline in every
single-agent workload and at most $0.699\%$ in the most parallel multi-agent cell.

\begin{figure}[htb]
\centering
\includegraphics[width=\columnwidth]{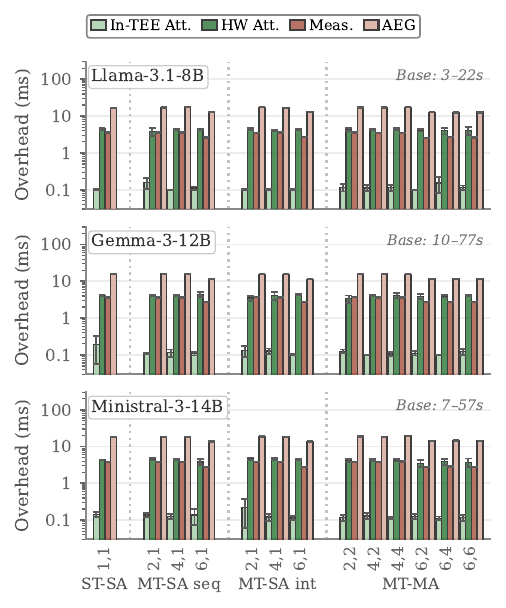}
\caption{\textbf{Stop Guard:} We report AEG generation, measurement, and attestation overheads across workloads and models (both in-TEE and hardware attestations) as a percentage of the baseline computation cost. We indicate the workloads as (\#task, \#agents). $y$-axis is log-scaled.}
\label{fig:interrupt}
\end{figure}

\end{itemize}
The guards are not meant to run in isolation, so we run all four on a single workload and report the end-to-end cost of the full monitor. 
We expect that composing them is free cryptographically, since every guard's evidence is folded into the same AEG and the AEG is attested once. 
We confirm this where the hardware attestation is $3.7-4.6$ms and the signature $\approx0.1$ms, exactly as for any single guard. \emph{AEG} is $16-30$ms, close to $\Gamma_{\mathrm{reg}}$'s alone because the judge dominates node count, and well below the sum of the four guards run separately. 
The full monitor costs $0.023-0.083\%$ of the baseline in every single-agent workload and at most $0.419\%$ in the most parallel multi-agent cell.

\begin{takeaway}
\textbf{Takeaway:} All guards incur $<$1\% overhead, and running all four guards together costs no more than running one.
\end{takeaway}

\subsection{Security Analysis}\label{sec:security}

Among the requirements for \method, \ref{compliance}, \ref{confidentiality}, and \ref{attestable} are security properties.
We argue that \method upholds these desired security properties under possible actions taken by \adv to bypass them, as defined in \S\ref{sec:problem}. 

\noindent\textbf{\ref{compliance} Compliance:} \adv cannot violate the specified regulation or induce unintended behavior (i.e., unauthorized tool or system calls). 
The APM binds the intended trajectory and guards to the user's prompt at generation time, and this binding is attested. \RefMon mediates every tool call the executor proposes, admitting an action only if it matches the expected tool call at its position in the trajectory and if it satisfies all applicable guards. Prompt injection can only cause harm by inducing the executor to propose an off-trajectory tool call or an unauthorized system call. The former is blocked with perfect recall because any position or parameter mismatch changes the hash measurement, while the latter is blocked by comparing every system call issued by the action against the per-prompt authorized set, derived from the toolkits actually used. The kernel sandbox provides defense-in-depth against unmediated system calls issued from inside a tool's implementation. Thus, when \adv manipulates only external sources, every deviation from the attested policy is either blocked at the tool interface by \RefMon or at the kernel layer by $\Gamma_{\mathrm{sys}}$, and the decision is recorded. 

\noindent\textbf{\ref{confidentiality} Confidentiality:} All components of \method execute inside the CVM and GPU TEEs. Intel TDX encrypts and integrity-protects CPU memory from the untrusted virtual machine monitor (VMM), while NVIDIA H100's confidential computing mode extends this protection to GPU memory and all data transferred between the CPU and GPU. Since \adv controls the host VMM, they can only observe encrypted memory. Model weights, prompts, intermediate outputs, and the APM remain in plaintext only inside the trusted boundary of \method. Verifiers receive only the AEG and its attestation, which contain no model weights. Thus, \adv learns nothing about the models beyond what is revealed by the AEG itself. 

\noindent\textbf{\ref{attestable} Attestability:} \adv may attempt to bypass the attestability by forging evidence sent to the verifier or by avoiding its actions being recorded in the evidence by \method. 
The AEG records every admitted and blocked action as a node with a measurement, all of which are used to produce a hash chain that commits to the content and order, along with a structure root that commits to the graph topology. A single Intel TDX quote and NVIDIA GPU token includes extending these measurements into the report data before they are signed, making the entire AEG tamper-evident. Any omission, reordering, modification, or inserted node changes either the chain head or the structure root, which the verifier can recompute offline to detect. The in-TEE signing key signs per-node evidence, and is certified by the same attestation. Because \RefMon is the only component that writes the AEG and invokes the attestor, and \RefMon itself runs inside the TEE, \adv cannot forge or alter the execution graph without detection. Any mismatch or invalid attestation indicates tampering and is detected by verifiers. 

\begin{takeaway}
\textbf{Takeaway:} \method provides compliance, confidentiality, and attestability despite the actions undertaken by \adv.
\end{takeaway}
\section{Discussion and Summary}\label{sec:discussions}

\noindent\textbf{Side Channel Attacks:} \method's security guarantees rely on the TEE's hardware isolation, but side channels remain a concern. Prior work has demonstrated that microarchitectural side channels (e.g., cache timing, page faults) can leak secrets from TEEs~\cite{teefail}. \adv controlling the host could exploit such channels to infer weights, prompt content, or the structure of the APM. 
Also, the executor's actions exhibit timing variations that depend on the trajectory taken, the tools invoked, and the data processed. 
Recent work shows that timing behavior of agentic workloads can be fingerprinted to recover task structure and tool calls~\cite{carlini2024remote,jeong2026network}. 

\method does not eliminate these channels, but requires a set of complementary approaches, such as constant time execution~\cite{cauligi2017fact,barthe2019formal}, cache partitioning~\cite{wang2007new,arm_cca}, or taint tracking of private data~\cite{blime,blackout}. 
We also note that the threat model of \method rules out attacks requiring physical access to the machine. 
Given recent attacks on TEEs that interpose DDR5 memory to leak secrets~\cite{teefail,batteringRAM,wiretap}, the security properties of \method would apply after manufacturers implement defenses against them.

\noindent\textbf{Support for Other AI Systems:} Although we focus on agentic systems, which increasingly underpin chatbot services, \method also supports standard chatbots. In this setting, the APM is replaced by a judge that evaluates input–output appropriateness, eliminating the need for $\Gamma_{\mathrm{sys}}$. Attestations capture inputs and outputs throughout the chat session~\cite{chantasantitam2026pal}.

\noindent\textbf{Trust in LLM Judge:} $\Gamma_{\mathrm{reg}}$ delegates compliance decisions to an LLM judge, whose accuracy is measured empirically rather than proven. This introduces a trust assumption: the judge's behavior must be stable across each of its operations. A malicious judge could admit non-compliant disclosures or block compliant ones, undermining the regulatory guarantee. Verifiers can audit the judge by inspecting its hash and evaluating it independently on a representative disclosure corpus. The judge's non-determinism is mitigated by using a fixed temperature and chain-of-thought prompting, but residual variance remains which is averaged across multiple runs. Future work includes replacing the judge with a formally verified checks with policies, or using multiple judges to reduce reliance on a single model.

\noindent\textbf{Agent Sandbox ``Escapes'':} Recent incidents have shown that agents can gain unauthorized access to the network (e.g., the OpenAI-Huggingface incident~\cite{openai2026huggingface}). 
\method can derive a per-prompt set of authorized system calls from the trajectory, including network calls only when required. 
\method can block network syscalls at the kernel level, providing least-privilege, attestable network containment.
APM exposes the authorized set, while the AEG records every permitted or blocked syscall. Verifiers can therefore confirm offline that (a) no network syscalls were permitted, (b) the kernel enforced this policy, and (c) no network access occurred or any attempted access was blocked and recorded.

\noindent\textbf{Other TEE Architectures:} Although \method is demonstrated with Intel TDX and NVIDIA H100, its design is technology-agnostic. APM, \RefMon, and AEG require only (1) a CVM providing memory confidentiality, integrity, and hardware-rooted remote attestation; (2) a GPU TEE providing equivalent guarantees for GPU memory and CPU--GPU communication and producing its own attestation; and (3) a Linux kernel supporting system-call filtering. These requirements can be met by other architectures, such as AMD SEV-SNP~\cite{amd-sev-snp} or ARM CCA~\cite{arm_cca}, with NVIDIA H100/Blackwell or equivalent GPUs. Thus, \method can be ported to any CVM/GPU TEE meeting the requirements.

\noindent\textbf{Summary:}
We present \method, the first design and implementation of a TEE-based regulatory sandbox. We identify key design requirements and incorporate them into \method, including the APM and the AEG that \RefMon compares to detect deviations from intended behavior, i.e., unauthorized tool and system calls. While prior works address individual compliance aspects, such as information flow control, policy compliance, and CFI, none systematically satisfy the requirements of regulatory sandboxes. Our evaluation demonstrates accurate compliance checking and low attestation overhead, making \method practical for running agents in the sandbox for monitoring by multiple verifiers prior to deployment.

\noindent\textbf{Broader Impact:} We address an immediate policy gap by proposing concrete mechanisms for regulatory sandboxes that preserve model confidentiality, support multi-verifier attestations, and enforce behavioral compliance of models. 
These mechanisms offer actionable tools for regulators and providers conducting pre-deployment oversight.
Although blocking off-trajectory but benign actions may reduce utility, we argue that blockage rates are an acceptable signal in sandbox settings, consistent with governance objectives.

\begin{arxiv}
\section*{Disclosure of AI Tool-Use}

\noindent We used Claude Code to assist with selected parts of the implementation, which the authors subsequently verified, and to help create Figure~\ref{fig:tree} and~\ref{fig:scalability} in TikZ. The authors independently developed the research ideas, methodology, experiments, and initial manuscript. We used ChatGPT selectively for language editing, primarily to correct grammar and conciseness. Thus, AI tools were used as assistants for limited implementation and editorial tasks.

\section*{Acknowledgments}
\noindent This work is supported by NSERC (Discovery grant) and the Canada CIFAR AI Chairs program.
Resources used in preparing this research were provided, in part, by the Province of Ontario, the Government of Canada through CIFAR, and companies sponsoring the Vector Institute.
\end{arxiv}

{\footnotesize
\bibliographystyle{IEEEtran}
\bibliography{paper}

@inproceedings{cauligi2017fact,
  title={{FaCT}: {A} flexible, constant-time programming language},
  author={Cauligi, Sunjay and Soeller, Gary and Brown, Fraser and Johannesmeyer, Brian and Huang, Yunlu and Jhala, Ranjit and Stefan, Deian},
  booktitle={IEEE Cybersecurity Development Conference (SecDev)},
  pages={69--76},
  year={2017}
}

@article{barthe2019formal,
  title={Formal verification of a constant-time preserving {C} compiler},
  author={Barthe, Gilles and Blazy, Sandrine and Gr{\'e}goire, Benjamin and Hutin, R{\'e}mi and Laporte, Vincent and Pichardie, David and Trieu, Alix},
  journal={ACM on Programming Languages (PACMPL)},
  volume={4},
  number={POPL},
  pages={1--30},
  year={2019}
}

@inproceedings{wang2007new,
  title={New cache designs for thwarting software cache-based side channel attacks},
  author={Wang, Zhenghong and Lee, Ruby B},
  booktitle={International Symposium on Computer Architecture (ISCA)},
  pages={494--505},
  year={2007}
}

@inproceedings{blackout,
  author       = {Hossam ElAtali and
                  Merve G{\"{u}}lmez and
                  Thomas Nyman and
                  N. Asokan},
  title        = {{BLACKOUT:} Data-Oblivious Computation with Blinded Capabilities},
  booktitle    = {ACM SIGSAC Conference on Computer and Communications Security (CCS)},
  pages        = {2039--2053},
  year         = {2025},
  url          = {https://doi.org/10.1145/3719027.3765169},
  doi          = {10.1145/3719027.3765169}
}

@inproceedings{blime,
  author       = {Hossam ElAtali and
                  Lachlan J. Gunn and
                  Hans Liljestrand and
                  N. Asokan},
  title        = {{BliMe}: Verifiably Secure Outsourced Computation with Hardware-Enforced
                  Taint Tracking},
  booktitle    = {Network and Distributed System Security Symposium (NDSS)},
  year         = {2024},
  url          = {https://www.ndss-symposium.org/ndss-paper/blime-verifiably-secure-outsourced-computation-with-hardware-enforced-taint-tracking/},
  bibsource    = {dblp computer science bibliography, https://dblp.org}
}

@inproceedings{teefail,
  title={{TEE.fail}: Breaking trusted execution environments via {DDR5} memory bus interposition},
  author={Chuang, Jalen and Seto, Alex and Berrios, Nicolas and Van Schaik, Stephan and Garman, Christina and Genkin, Daniel},
  booktitle={IEEE Symposium on Security and Privacy (S\&P)},
  pages={4408--4426},
  year={2026}
}

@inproceedings{li2025privacy,
  title={Privacy checklist: Privacy violation detection grounding on contextual integrity theory},
  author={Li, Haoran and Fan, Wei and Chen, Yulin and Jiayang, Cheng and Chu, Tianshu and Zhou, Xuebing and Hu, Peizhao and Song, Yangqiu},
  booktitle={Conference of the Nations of the Americas Chapter of the Association for Computational Linguistics (NAACL)},
  pages={1748--1766},
  year={2025}
}

@inproceedings{fan2024goldcoin,
  title={{GoldCoin}: Grounding large language models in privacy laws via contextual integrity theory},
  author={Fan, Wei and Li, Haoran and Deng, Zheye and Wang, Weiqi and Song, Yangqiu},
  booktitle={Conference on Empirical Methods in Natural Language Processing (EMNLP)},
  pages={3321--3343},
  year={2024}
}

@inproceedings{li-contextlens,
    title = "{C}ontext{L}ens: Modeling Imperfect Privacy and Safety Context for Legal Compliance",
    author = "Li, Haoran  and
      Chen, Yulin  and
      Jing, Huihao  and
      Hu, Wenbin  and
      Li, Tsz Ho  and
      Lou, Chanhou  and
      Tsang, Hong Ting  and
      Han, Sirui  and
      Song, Yangqiu",
    booktitle = "Annual Meeting of the Association for Computational Linguistics (ACL)",
    month = jul,
    year = "2026",
    address = "San Diego, California, United States",
    url = "https://aclanthology.org/2026.acl-long.294/",
    doi = "10.18653/v1/2026.acl-long.294",
    pages = "6503--6518",
    ISBN = "979-8-89176-390-6"
}

@article{wang2026benchmarking,
  title={Benchmarking Confidential {GPU} Inference on {NVIDIA} {H100} under Intel {TDX}},
  author={Wang, Wei and others},
  journal={arXiv preprint arXiv:2607.19353},
  year={2026}
}

@article{zhu2024confidential,
  title={Confidential computing on nvidia hopper gpus: {A} performance benchmark study},
  author={Zhu, Jianwei and others},
  journal={arXiv preprint arXiv:2409.03992},
  year={2024}
}

@inproceedings{chrapek2025confidential,
  title={Confidential llm inference: Performance and cost across cpu and gpu tees},
  author={Chrapek, Marcin and others},
  booktitle={IEEE International Symposium on Workload Characterization (IISWC)},
  pages={84--98},
  year={2025}
}

@inproceedings{batteringRAM,
  title={Battering {RAM}: Low-cost interposer attacks on confidential computing via dynamic memory aliasing},
  author={De Meulemeester, Jesse and Oswald, David and Verbauwhede, Ingrid and Van Bulck, Jo},
  booktitle={IEEE Symposium on Security and Privacy (S\&P)},
  pages={4391--4407},
  year={2026}
}

@inproceedings{wiretap,
  title={Wiretap: Breaking server sgx via dram bus interposition},
  author={Seto, Alex and Duran, Oytun Kuday and Amer, Samy and Chuang, Jalen and van Schaik, Stephan and Genkin, Daniel and Garman, Christina},
  booktitle={ACM SIGSAC Conference on Computer and Communications Security (CCS)},
  pages={708--722},
  year={2025}
}

@online{openai2026huggingface,
  author  = {{OpenAI}},
  title   = {{OpenAI} and Hugging Face Partner to Address Security Incident During Model Evaluation},
  year    = {2026},
  month   = jul,
  day     = {21},
  url     = {https://openai.com/index/hugging-face-model-evaluation-security-incident/},
  note    = {Updated July 29, 2026}
}

@article{
siddiqui2025permissive,
title={Permissive Information-Flow Analysis for Large Language Models},
author={Shoaib Ahmed Siddiqui and others},
journal={Transactions on Machine Learning Research (TMLR)},
issn={2835-8856},
year={2025},
url={https://openreview.net/forum?id=ufYRO8y3mr},
note={}
}

@inproceedings{
kolluri2026optimizing,
title={Optimizing Agent Planning for Security and Autonomy},
author={Aashish Kolluri and others},
booktitle={International Conference on Learning Representations (ICLR)},
year={2026},
url={https://openreview.net/forum?id=g0aVCDY3gS}
}

@article{costa2025securing,
  title={Securing {AI} agents with information-flow control},
  author={Costa, Manuel and others},
  journal={arXiv preprint arXiv:2505.23643},
  year={2025}
}

@article{kim2025prompt,
  title={Prompt flow integrity to prevent privilege escalation in llm agents},
  author={Kim, Juhee and others},
  journal={arXiv preprint arXiv:2503.15547},
  year={2025}
}

@inproceedings{tsai2025contextual,
  title={Contextual agent security: {A} policy for every purpose},
  author={Tsai, Lillian and Bagdasarian, Eugene},
  booktitle={Workshop on Hot Topics in Operating Systems (HotOS)},
  pages={8--17},
  year={2025}
}

@article{Wang2026AligningPW,
  title={Aligning Provenance with Authorization: {A} Dual-Graph Defense for {LLM} Agents},
  author={Peiran Wang and others},
  journal={arXiv preprint arXiv:2605.26497},
  year={2026}
}

@article{kravchenko2026agentic,
  title={Agentic Permissions Policy Algebra for Taint Confinement in {LLM} Agents},
  author={Kravchenko, Arseny and others},
  journal={arXiv preprint arXiv:2607.24625},
  year={2026}
}

@inproceedings{jeong2026network,
  title={$\{$Network-Level$\}$ Prompt and Trait Leakage in Local Research Agents},
  author={Jeong, Hyejun and Teymoorianfard, Mohammadreza and Kumar, Abhinav and Houmansadr, Amir and Bagdasarian, Eugene},
  booktitle={USENIX Security Symposium (USENIX Security)},
  pages={6485--6504},
  year={2026}
}

@article{carlini2024remote,
  title={Remote timing attacks on efficient language model inference},
  author={Carlini, Nicholas and Nasr, Milad},
  journal={arXiv preprint arXiv:2410.17175},
  year={2024}
}

@article{zhong2025rtbas,
  title={Rtbas: Defending llm agents against prompt injection and privacy leakage},
  author={Zhong, Peter Yong and others},
  journal={arXiv preprint arXiv:2502.08966},
  year={2025}
}

@article{alam2026prompt,
  title={Prompt Control-Flow Integrity: {A} Priority-Aware Runtime Defense Against Prompt Injection in {LLM} Systems},
  author={Alam, Md Takrim Ul and others},
  journal={arXiv preprint arXiv:2603.18433},
  year={2026}
}

@article{palumbo2026formal,
  title={Formal policy enforcement for real-world agentic systems},
  author={Palumbo, Nils and others},
  journal={arXiv preprint arXiv:2602.16708},
  year={2026}
}

@article{solko2026efficient,
  title={Efficient and Sound Probabilistic Verification for {AI} Agents},
  author={Solko-Breslin, Alaia and others},
  journal={arXiv preprint arXiv:2606.20510},
  year={2026}
}

@article{meng2025cellmate,
  title={{CellMate}: Sandboxing browser {AI} agents},
  author={Meng, Luoxi and others},
  journal={arXiv preprint arXiv:2512.12594},
  year={2025}
}

@inproceedings{
balunovic2024ai,
title={{AI} Agents with Formal Security Guarantees},
author={Mislav Balunovic and others},
booktitle={ICML Workshop on Next Generation of AI Safety},
year={2024},
url={https://openreview.net/forum?id=c6jNHPksiZ}
}

@inproceedings{waseem2026engineering,
  title={Engineering a Governance-Aware {AI} Sandbox: Design, Implementation, and Lessons Learned},
  author={Waseem, Muhammad and others},
  booktitle={International Conference on Evaluation and Assessment in Software Engineering (EASE)},
  year={2026}
}

@online{nanda2026nest,
  author  = {{NANDA}},
  title   = {{NEST}: {NANDA} Exchange Sandbox \& {TestNet}},
  year    = {2026},
  url     = {https://nest.projectnanda.org/},
  note    = {NANDA Exchange Sandbox and TestNet}
}

@article{sidik2026beyond,
  title={Beyond static sandboxing: Learned capability governance for autonomous ai agents},
  author={Sidik, Bronislav and Rokach, Lior},
  journal={arXiv preprint arXiv:2604.11839},
  year={2026}
}

@article{buscemi2025sandboxes,
  author  = {Buscemi, Alessio and others},
  title   = {Towards Sandboxes for the Internet of Agents},
  journal = {Social Science Research Network (SSRN)},
  year    = {2025},
  month   = nov,
  day     = {21},
  doi     = {10.2139/ssrn.5801322},
  url     = {https://papers.ssrn.com/sol3/papers.cfm?abstract_id=5801322}
}

@article{piao2025agentbay,
  title={Agentbay: {A} hybrid interaction sandbox for seamless human-ai intervention in agentic systems},
  author={Piao, Yun and others},
  journal={arXiv preprint arXiv:2512.04367},
  year={2025}
}

@online{googlecloud2026agentsandbox,
  author  = {{Google Cloud}},
  title   = {About {GKE} Agent Sandbox},
  year    = {2026},
  month   = aug,
  day     = {20},
  url     = {https://docs.cloud.google.com/kubernetes-engine/docs/concepts/machine-learning/agent-sandbox},
  note    = {Google Cloud Documentation}
}

@inproceedings{ahern2025operationalising,
  title={Operationalising {AI} regulatory sandboxes under the {EU} {AI} Act: The triple challenge of capacity, coordination and attractiveness to providers},
  author={Ahern, Deirdre},
  booktitle={Cambridge Forum on AI: Law and Governance},
  volume={1},
  pages={e35},
  year={2025}
}

@inproceedings {TiwariIFC,
author = {Trishita Tiwari and others},
title = {Information Flow Control in Machine Learning through Modular Model Architecture},
booktitle = {USENIX Security Symposium (USENIX Security)},
year = {2024},
pages = {6921--6938},
url = {https://www.usenix.org/conference/usenixsecurity24/presentation/tiwari}
}

@article{cai2026ghost,
  title={Ghost in the agent: Redefining information flow tracking for llm agents},
  author={Cai, Yuandao and others},
  journal={arXiv preprint arXiv:2604.23374},
  year={2026}
}

@article{debenedetti2025defeating,
  title={Defeating prompt injections by design},
  author={Debenedetti, Edoardo and others},
  journal={arXiv preprint arXiv:2503.18813},
  year={2025}
}

@article{wang2025agentarmor,
  title={{AgentArmor}: Enforcing program analysis on agent runtime trace to defend against prompt injection},
  author={Wang, Peiran and others},
  journal={arXiv preprint arXiv:2508.01249},
  year={2025}
}

@article{beurer2025design,
  title={Design patterns for securing llm agents against prompt injections},
  author={Beurer-Kellner, Luca and others},
  journal={arXiv preprint arXiv:2506.08837},
  year={2025}
}

@inproceedings{wu2024isolategpt,
  title={Isolategpt: An execution isolation architecture for llm-based agentic systems},
  author={Wu, Yuhao and others},
  booktitle={Network and Distributed System Security Symposium (NDSS)},
  year={2025}
}

@article{del2025architecting,
  title={Architecting resilient llm agents: {A} guide to secure plan-then-execute implementations},
  author={Del Rosario, Ron F and others},
  journal={arXiv preprint arXiv:2509.08646},
  year={2025}
}

@inproceedings{grigor2026vet,
  title={Vet your agent: Towards host-independent autonomy via verifiable execution traces},
  author={Grigor, Artem and others},
  booktitle={ACM Asia Conference on Computer and Communications Security (AsiaCCS)},
  pages={820--834},
  year={2026}
}

@inproceedings{abdollahi2026agentee,
  title={{AgenTEE}: Confidential {LLM} Agent Execution on Edge Devices},
  author={Abdollahi, Sina and others},
  booktitle={European Workshop on Machine Learning and Systems (EuroMLSys)},
  pages={473--480},
  year={2026}
}

@article{bodea2025trusted,
  title={Trusted {AI} agents in the cloud},
  author={Bodea, Teofil and others},
  journal={arXiv preprint arXiv:2512.05951},
  year={2025}
}

@article{villa2026prismata,
  title={Prismata: Confining Cross-Site Prompt Injection in Web Agents},
  author={Villa, Corban and others},
  journal={arXiv preprint arXiv:2607.08147},
  year={2026}
}

@article{piet2026web,
  title={Web Agents Should Adopt the Plan-Then-Execute Paradigm},
  author={Piet, Julien and others},
  journal={arXiv preprint arXiv:2605.14290},
  year={2026}
}

@article{ngong2026agentscope,
  title={{AgentSCOPE}: Evaluating contextual privacy across agentic workflows},
  author={Ngong, Ivoline C and others},
  journal={arXiv preprint arXiv:2603.04902},
  year={2026}
}

@inproceedings{bagdasarian2024airgapagent,
  title={Airgapagent: Protecting privacy-conscious conversational agents},
  author={Bagdasarian, Eugene and others},
  booktitle={ACM SIGSAC Conference on Computer and Communications Security (CCS)},
  pages={3868--3882},
  year={2024}
}

@article{
ghalebikesabi2025privacy,
title={Privacy Awareness for Information-Sharing Assistants: {A} Case-study on Form-filling with Contextual Integrity},
author={Sahra Ghalebikesabi and others},
journal={Transactions on Machine Learning Research (TMLR)},
issn={2835-8856},
year={2025},
url={https://openreview.net/forum?id=l9rATNBB8Y}
}

@article{foerster2026camels,
  title={Camels can use computers too: System-level security for computer use agents},
  author={Foerster, Hanna and others},
  journal={arXiv preprint arXiv:2601.09923},
  year={2026}
}

@inproceedings{
jha2026breaking,
title={Breaking and Fixing Defenses Against Control Flow Hijacking in Multi-Agent Systems},
author={Rishi Dev Jha and others},
booktitle={International Conference on Learning Representations (ICLR)},
year={2026},
url={https://openreview.net/forum?id=PNU9Rj5RDQ}
}

@article{li2026drift,
  title={{DRIFT}: Dynamic rule-based defense with injection isolation for securing {LLM} agents},
  author={Li, Hao and others},
  journal={Advances in Neural Information Processing Systems (NeurIPS)},
  volume={38},
  pages={83262--83290},
  year={2026}
}

@inproceedings{wang2025agentspec,
  title={Agentspec: Customizable runtime enforcement for safe and reliable llm agents},
  author={Wang, Haoyu and others},
  booktitle={IEEE/ACM International Conference on Software Engineering (ICSE)},
  year={2026}
}

@inproceedings{chen2025shieldagent,
  title={Shieldagent: Shielding agents via verifiable safety policy reasoning},
  author={Chen, Zhaorun and others},
  booktitle={International Conference on Machine Learning (ICML)},
  pages={8313--8344},
  year={2025}
}

@inproceedings{hua2024trustagent,
  title={Trustagent: Towards safe and trustworthy llm-based agents},
  author={Hua, Wenyue and others},
  booktitle={Findings of the Association for Computational Linguistics (EMNLP Findings)},
  pages={10000--10016},
  year={2024}
}

@inproceedings{jiang2026chaincaps,
  title={{ChainCaps}: Composition-Safe Tool-Using Agents via Monotonic Capability Attenuation},
  author={Jiang, Xiaochong and others},
  booktitle={ICML Workshop on Agents in the Wild: Safety, Security, and Beyond (AIWILD)},
  year={2026}
}

@article{abaev2026agentguardian,
  title={Agentguardian: Learning access control policies to govern ai agent behavior},
  author={Abaev, Nadya and others},
  journal={arXiv preprint arXiv:2601.10440},
  year={2026}
}

@article{chantasantitam2026pal,
  title={{PAL}$^{*}${M}: Property attestation for large generative models},
  author={Chantasantitam, Prach and others},
  journal={arXiv preprint arXiv:2601.16199},
  year={2026}
}

@inproceedings{duddu2024laminator,
  title={Laminator: Verifiable {ML} property cards using hardware-assisted attestations},
  author={Duddu, Vasisht and others},
  booktitle={ACM Conference on Data and Application Security and Privacy (CODASPY)},
  pages={317--328},
  year={2024}
}

@inproceedings{greshake2023not,
  title={Not what you've signed up for: Compromising real-world llm-integrated applications with indirect prompt injection},
  author={Greshake, Kai and others},
  booktitle={ACM Workshop on Artificial Intelligence and Security (AISec)},
  pages={79--90},
  year={2023}
}

@inproceedings{
debenedetti2024agentdojo,
title={{AgentDojo}: {A} Dynamic Environment to Evaluate Prompt Injection Attacks and Defenses for {LLM} Agents},
author={Edoardo Debenedetti and others},
booktitle={Conference on Neural Information Processing Systems Datasets and Benchmarks Track (NeurIPS D\&B)},
year={2024},
url={https://openreview.net/forum?id=m1YYAQjO3w}
}

@inproceedings{zhan2024injecagent,
  title={Injecagent: Benchmarking indirect prompt injections in tool-integrated large language model agents},
  author={Zhan, Qiusi and others},
  booktitle={Findings of the Association for Computational Linguistics (ACL Findings)},
  pages={10471--10506},
  year={2024}
}

@inproceedings{usenixBenchmark,
author = {Yupei Liu and others},
title = {Formalizing and Benchmarking Prompt Injection Attacks and Defenses},
booktitle = {USENIX Security Symposium (USENIX Security)},
year = {2024},
pages = {1831--1847},
url = {https://www.usenix.org/conference/usenixsecurity24/presentation/liu-yupei}
}

@article{Wei2026AgenticRF,
  title={Agentic Reasoning for Large Language Models},
  author={Tianxin Wei and others},
  journal={arXiv preprint arXiv:2502.04644},
  year={2026},
  volume={abs/2601.12538},
  url={https://arxiv.org/abs/2601.12538}
}

@online{euaiactarticle57,
  title        = {Article 57: {AI} Regulatory Sandboxes},
  author       = {{EU Artificial Intelligence Act}},
  year         = {2026},
  url          = {https://artificialintelligenceact.eu/article/57/},
  urldate      = {2026-09-06}
}

@online{euaiactsandbox_overview,
  title        = {{AI} Regulatory Sandbox Approaches: {EU} Member State Overview},
  author       = {{EU Artificial Intelligence Act}},
  year         = {2026},
  url          = {https://artificialintelligenceact.eu/ai-regulatory-sandbox-approaches-eu-member-state-overview/},
  urldate      = {2026-09-06}
}

@online{whitecasespain,
  title        = {{AI} Watch: Global Regulatory Tracker --- Spain},
  author       = {{White \& Case LLP}},
  year         = {2026},
  url          = {https://www.whitecase.com/insight-our-thinking/ai-watch-global-regulatory-tracker-spain},
  urldate      = {2026-09-06}
}

@online{ibasandboxes,
  title        = {{AI} and Regulatory Sandboxes},
  author       = {{International Bar Association}},
  year         = {2026},
  url          = {https://www.ibanet.org/AI-and-regulatory-sandboxes},
  urldate      = {2026-09-06}
}

@online{iappsingapore,
  title        = {Singapore Offers Potential Blueprint for {AI} Sandbox Initiatives},
  author       = {{International Association of Privacy Professionals}},
  year         = {2026},
  url          = {https://iapp.org/news/a/as-ai-sandboxes-are-considered-singapore-offers-a-blueprint},
  urldate      = {2026-09-06}
}

@online{fpfsandboxes,
  title        = {Balancing Innovation and Oversight: Regulatory Sandboxes as a Tool for {AI} Governance},
  author       = {{Future of Privacy Forum}},
  year         = {2025},
  url          = {https://fpf.org/blog/balancing-innovation-and-oversight-regulatory-sandboxes-as-a-tool-for-ai-governance/},
  urldate      = {2026-09-06}
}

@online{aisafetysandbox,
  title        = {{AI} Regulatory Sandboxes: Testing {AI} Innovation Under Supervision},
  author       = {{AI Safety Directory}},
  year         = {2026},
  url          = {https://aisecurityandsafety.org/en/guides/ai-regulatory-sandbox/},
  urldate      = {2026-09-06}
}

@online{mideastuae,
  title        = {{UAE} {AI} Regulation 2026: Federal Framework Explained},
  author       = {{The Middle East Insider}},
  year         = {2026},
  url          = {https://themiddleeastinsider.com/2026/04/25/uae-ai-regulation-2026/},
  urldate      = {2026-09-06}
}

@online{iappjurisdictions,
  title        = {How Different Jurisdictions Approach {AI} Regulatory Sandboxes},
  author       = {{International Association of Privacy Professionals}},
  year         = {2026},
  url          = {https://iapp.org/news/a/how-different-jurisdictions-approach-ai-regulatory-sandboxes},
  urldate      = {2026-09-06}
}

@article{shi2025progent,
  title={Progent: Securing {AI} agents with privilege control},
  author={Shi, Tianneng and others},
  journal={arXiv preprint arXiv:2504.11703},
  year={2025}
}

@misc{azure_sandbox,
  author       = {{Microsoft}},
  title        = {Azure Sandbox},
  year         = {2026},
  howpublished = {\url{https://learn.microsoft.com/en-us/azure/architecture/guide/azure-sandbox/azure-sandbox}},
  note         = {Microsoft Learn, Azure Architecture Center. Accessed September 10, 2026}
}

@article{shu2016study,
  title={{A} study of security isolation techniques},
  author={Shu, Rui and others},
  journal={ACM Computing Surveys (CSUR)},
  volume={49},
  number={3},
  pages={1--37},
  year={2016}
}

@inproceedings{minibox,
  title={$\{${MiniBox}$\}$: {A} $\{$Two-Way$\}$ Sandbox for x86 Native Code},
  author={Li, Yanlin and others},
  booktitle={USENIX Annual Technical Conference (USENIX ATC)},
  pages={409--420},
  year={2014}
}

@inproceedings{liang2003isolated,
  title={Isolated program execution: An application transparent approach for executing untrusted programs},
  author={Venkatakrishnan, VN and Sekar, R},
  booktitle={Annual Computer Security Applications Conference (ACSAC)},
  pages={182--191},
  year={2003}
}

@inproceedings{castro2006securing,
  title={Securing software by enforcing data-flow integrity},
  author={Castro, Miguel and others},
  booktitle={USENIX Symposium on Operating Systems Design and Implementation (OSDI)},
  pages={147--160},
  year={2006}
}

@inproceedings{song2016enforcing,
  title={Enforcing Kernel Security Invariants with Data Flow Integrity.},
  author={Song, Chengyu and others},
  booktitle={Network and Distributed System Security Symposium (NDSS)},
  year={2016}
}

@article{buccafurri2024mqtt,
  title={{MQTT}-{I}: Achieving end-to-end data flow integrity in {MQTT}},
  author={Buccafurri, Francesco and others},
  journal={IEEE Transactions on Dependable and Secure Computing (TDSC)},
  volume={21},
  number={5},
  pages={4717--4734},
  year={2024}
}

@article{chen2022finedift,
  title={{FineDIFT}: Fine-grained dynamic information flow tracking for data-flow integrity using coprocessor},
  author={Chen, Kejun and others},
  journal={IEEE Transactions on Information Forensics and Security (TIFS)},
  volume={17},
  pages={559--573},
  year={2022}
}

@inproceedings{hdfi,
  title={{HDFI}: Hardware-assisted data-flow isolation},
  author={Song, Chengyu and others},
  booktitle={IEEE Symposium on Security and Privacy (S\&P)},
  pages={1--17},
  year={2016}
}

@inproceedings{datashield,
  title={Datashield: Configurable data confidentiality and integrity},
  author={Carr, Scott A and Payer, Mathias},
  booktitle={ACM Asia Conference on Computer and Communications Security (AsiaCCS)},
  pages={193--204},
  year={2017}
}

@inproceedings{pac-it-up,
  title={{PAC} it up: Towards Pointer Integrity using {ARM} Pointer Authentication.},
  author={Liljestrand, Hans and others},
  booktitle={USENIX Security Symposium (USENIX Security)},
  pages={177--194},
  year={2019}
}

@inproceedings{pal,
  title={{In-Kernel} {Control-Flow} Integrity on Commodity {OSes} using {ARM} Pointer Authentication},
  author={Yoo, Sungbae and others},
  booktitle={USENIX Security Symposium (USENIX Security)},
  pages={89--106},
  year={2022}
}

@inproceedings{pacstack,
  title={{PACStack}: an Authenticated Call Stack.},
  author={Liljestrand, Hans and others},
  booktitle={USENIX Security Symposium (USENIX Security)},
  pages={357--374},
  year={2021}
}

@inproceedings{kuzhiyelil2020towards,
  title={Towards transparent control-flow integrity in safety-critical systems},
  author={Kuzhiyelil, Don and others},
  booktitle={International Conference on Information Security (ISC)},
  pages={290--311},
  year={2020}
}

@inproceedings{sherloc,
  title={{SHERLOC}: Secure and Holistic Control-Flow Violation Detection on Embedded Systems},
  author={Tan, Xi and Zhao, Ziming},
  booktitle={ACM SIGSAC Conference on Computer and Communications Security (CCS)},
  pages={1332--1346},
  year={2023}
}

@inproceedings{flowguard,
  title={Transparent and efficient cfi enforcement with intel processor trace},
  author={Liu, Yutao and others},
  booktitle={IEEE International Symposium on High-Performance Computer Architecture (HPCA)},
  pages={529--540},
  year={2017}
}

@inproceedings{typesqueezer,
  title={{TypeSqueezer}: When Static Recovery of Function Signatures for Binary Executables Meets Dynamic Analysis},
  author={Lin, Ziyi and others},
  booktitle={ACM SIGSAC Conference on Computer and Communications Security (CCS)},
  pages={2725--2739},
  year={2023}
}

@inproceedings{hek-cfi,
  title={Beyond the Edges of Kernel Control-Flow Hijacking Protection with {HEK}-{CFI}},
  author={Maar, Lukas and others},
  booktitle={ACM Asia Conference on Computer and Communications Security (AsiaCCS)},
  year={2024}
}

@inproceedings{sok_cfa_cfi,
  title={{SoK}: Integrity, Attestation, and Auditing of Program Execution},
  author={Ammar, Mahmoud and others},
  booktitle={IEEE Symposium on Security and Privacy (S\&P)},
  pages={77--77},
  year={2024}
}

@MISC{tdx,
  title={{I}ntel {Trust Domain Extensions} ({Intel} {TDX})},
  author={Intel},
  howpublished={\url{https://www.intel.com/content/www/us/en/developer/tools/trust-domain-extensions/overview.html}},
  year=2023 
}

@MISC{amd-sev-snp,
  title={{AMD} Secure Encrypted Virtualization ({SEV})},
  author={AMD},
  howpublished={\url{https://www.amd.com/en/developer/sev.html}},
  year=2020 
}

@article{h100,
  title={{NVIDIA Hopper H100 GPU: Scaling Performance}},
  author={Choquette, Jack},
  journal={IEEE Micro},
  volume={43},
  number={3},
  pages={9--17},
  year={2023}
}

@misc{arm_cca,
    title={Confidential Compute Architecture},
    author={ARM},
    howpubished={\url{https://www.arm.com/architecture/security-features/arm-confidential-compute-architecture}},
    year={[Online] Accessed September 2026}
}

@misc{eu_ai_act_multi_agent,
  author       = {Pieter van Schalkwyk},
  title        = {The {EU} {AI} Act now treats multi-agent systems as one system},
  year         = {2026},
  month        = may,
  howpublished = {\url{https://www.linkedin.com/pulse/eu-ai-act-now-treats-multi-agent-systems-one-system-van-schalkwyk-ondmc/}}
}
}

\appendices

\section{APM Example}\label{app:example}

\begin{figure}[!h]
\centering
\begin{lstlisting}[style=apm]
// Trajectory $\traj$: permitted tool calls, in order
trajectory: [
  { pos: 0, tool: Terminal.Execute,
    params: { command: "summarize report.csv" } },
  { pos: 1, tool: Gmail.SendEmail,  // disclosure
    params: { to: "auditor@acme.com" } } ]
// each tool call stores its hash $\mathrm{H}(e_j)$

// $\Gamma_{\mathrm{reg}}$: the governing regulation
{ name: regulation, scope: DISCLOSURE_TOOLS,
  predicate: "$\varphi(\act)$ permitted under R",
  spec: { regulations: [gdpr, hipaa] } }

// $\Gamma_{\mathrm{sys}}$: authorized system calls
{ name: os, scope: syscall,
  predicate: "all system calls of $\act$ in Auth",
  spec: {
    syscalls: [openat, read, write,  // runtime
               execve, fork,       // + Terminal
               connect, socket],   // + Gmail
    paths:   [/lib, /proc, /usr],
    rlimits: { NPROC: 1, FSIZE: 64MB } } }

// $\Gamma_{\mathrm{int}}$: budget and stop signal
{ name: interrupt, scope: continuation,
  predicate: "no trigger in $\sigma$",
  spec: { budget: { turns: 8 }, stop: external } }
\end{lstlisting}
\caption{\textbf{APM Example:} For tasks involving \texttt{Terminal} and \texttt{Gmail}, $\traj$ lists permitted tool calls in order, each with hash $\mathrm{H}(e_j)$ for $\Gamma_{\mathrm{traj}}$ to verify. Multi-agent runs include one branch per sub-agent. Each guard defines the check (\texttt{name}), covered actions (\texttt{scope}), condition (\texttt{predicate}), and ground truth (\texttt{spec}), which \RefMon independently evaluates using a static predicate or language-model oracle. \texttt{execve}/\texttt{socket} entries are included only when required by the trajectory's tools.}
\label{fig:apm-guards}
\end{figure}

\section{Attestation Definitions}\label{app:attestations}

\method attests every step of a run: the APM the planner derives, each model call, each decision \RefMon makes, each stop, each sub-task dispatch, and the AEG the run leaves behind.
All of them follow the same template: the \textbf{input} names the values the attestation binds, the \textbf{computation} is what runs inside the TD, the \textbf{output} is the signed statement a verifier receives, and the \textbf{assertion} is what that verifier may conclude once the signature checks out.

\noindent\textbf{\underline{APM Generation:}} This binds the policy to the inputs it was derived from, so that a policy cannot be swapped for a more permissive one later.
It takes the clean prompt $p$, the system-call manifest $C$, the applicable regulations $R$, and the stop condition, and outputs the APM $(\traj,\Gamma)$ together with a signature over the hashes of all of them.
A verifier concludes that the enforced trajectory came from the user's prompt alone and that the authorized set $\mathrm{Auth}$ covers the tools that trajectory require (\Cref{att:apm}).

\begin{attdesc}[htb]
\caption{APM Generation}\label{att:apm}
\textbf{Input:} prompt $p$, system-call manifest $C$, regulations $R$, stop condition $\mathrm{stop}$\;
\textbf{Computation:} planner derives APM $(\traj,\Gamma) \leftarrow \mathsf{Planner}(p, C, R, \mathrm{stop})$; $\mathrm{Auth}$ is union of system calls that tools on $\traj$ declare in $C$\;
\textbf{Output:} APM, $\signatt\bigl(\text{`APMAtt'}, \mathrm{H}(p, C, R, \traj, \Gamma)\bigr)$\;
\textbf{Assertion:} $\traj$ derives from $p$ alone, and $\mathrm{Auth}$ holds only what manifests of $\traj$'s tools require\;
\end{attdesc}

\noindent\textbf{\underline{Inference Binding:}} This binds a model inference to the weights that produced it, so that no model in the pipeline, planner, executor, or judge, can be silently substituted for another one.
It takes the model's weights, its input, and its output, and outputs a \texttt{proof\_of\_inference} recorded on the node that used the inference.
A verifier concludes that the recorded output was produced by the claimed model on the recorded input (\Cref{att:inf}).

\begin{attdesc}[htb]
\caption{Inference Binding}\label{att:inf}
\textbf{Input:} model weights $w$, input $x$ (prompt and context), output $y$\;
\textbf{Computation:} model uses inference logic spanning a TD and confidential-computing GPU (producing $GPU_{Att}$); \method produces $GPU_{Att}$ and hashes $w$ once per run as \texttt{weights\_hash}, and each $(x,y)$ pair per call\;
\textbf{Output:} $\signatt\bigl(\text{`InfAtt'}, \mathrm{H}(w, x, y), GPU_{Att}\bigr)$, recorded as \texttt{proof\_of\_inference}\;
\textbf{Assertion:} $y$ came from model with \texttt{weights\_hash} $\mathrm{H}(w)$, on input $x$, inside TD paired to GPU that produces $GPU_{Att}$\;
\end{attdesc}

\noindent\textbf{\underline{Reference Monitor Check:}} This binds \RefMon's decision to the action and the guards, so that an admitted action cannot later be presented as unchecked, nor a blocked one as never attempted.
It takes the proposed action $\act$ at position $j$, the expected tool call $e_j$, and the guards $\Gamma_{\mid \act}$ that cover $\act$, and outputs the verdict together with the measurement of the \texttt{check} node that records it.
A verifier concludes that $\act$ was evaluated against every applicable guard, and that the tool ran only if all of them passed (\Cref{att:check}).

\begin{attdesc}[htb]
\caption{Reference Monitor Check}\label{att:check}
\textbf{Input:} proposed action $\act$ at position $j$, expected tool call $e_j$, applicable guards $\Gamma_{\mid \act} \subseteq \Gamma$, run state $\sigma$\;
\textbf{Computation:} \RefMon evaluates every guard in $\Gamma_{\mid \act}$ and emits $v \leftarrow \textsc{allow}$ if all hold, else $v \leftarrow \textsc{block}$; tool executes only when $v = \textsc{allow}$, and verdict is committed as \texttt{check} node with measurement $m_i$\;
\textbf{Output:} $\signatt\bigl(\text{`ChkAtt'}, \mathrm{H}(\act, e_j, \Gamma_{\mid \act}, v, m_i)\bigr)$\;
\textbf{Assertion:} $\act$ passed through \RefMon and was checked against every guard in $\Gamma_{\mid \act}$ before any side effect, and tool executed iff $v = \textsc{allow}$\;
\end{attdesc}

\noindent\textbf{\underline{Regulation Judgment:}} $\Gamma_{\mathrm{reg}}$ uses oracle rather than a decidable predicate (\S\ref{sec:refmonitor}), so its attestation additionally binds \emph{which} judge decided, what it was shown, and what it answered.
It takes the action tuple $\varphi=\derive(\act)$, regulation $R$, and the judge $J$, and outputs the label the judge returned.
A verifier concludes that the label was produced by that judge on that tuple under $R$. Its \emph{correctness} is an empirical property of $J$ (\S\ref{sec:evaluation}; \Cref{att:reg}).

\begin{attdesc}[htb]
\caption{Regulation Judgment}\label{att:reg}
\textbf{Input:} action tuple $\varphi = \derive(\act)$, regulation text $R$, fixed judge $J$ with weights $w_J$, fixed prompt template with $k$-shot examples\;
\textbf{Computation:} \RefMon queries $J$ on $(\varphi, R)$ under fixed template and obtains label $\ell$: compliant, non-compliant, or not-applicable; action is blocked iff $\ell$ is non-compliant\;
\textbf{Output:} $\signatt\bigl(\text{`RegAtt'}, \mathrm{H}(\varphi, R, w_J, \textit{template}, \ell)\bigr)$, with \texttt{proof\_of\_inference} of judge call (\Cref{att:inf})\;
\textbf{Assertion:} $\ell$ came from fixed judge $J$ on $\varphi$ under $R$, and action was blocked iff $\ell$ is non-compliant\;
\end{attdesc}

\noindent\textbf{\underline{Interrupt:}} This binds a stop to the point at which it happened, so that a truncated run cannot be mistaken for a tampered one, nor a tampered one presented as merely interrupted.
It takes the run state at the stop and the trigger that fired, and outputs the \texttt{interrupt} node that terminates the branch.
A verifier concludes that the run stopped at that position, that the pending action never executed, and that nothing follows it in that branch (\Cref{att:int}).

\begin{attdesc}[htb]
\caption{Interrupt}\label{att:int}
\textbf{Input:} run state $\sigma$ at stop, trigger that fired (timeout, resource cap, or external stop), position $j$ of pending action\;
\textbf{Computation:} \RefMon neither executes nor commits action at $j$, and commits \texttt{interrupt} node with measurement $m_i$ as terminal event of branch\;
\textbf{Output:} $\signatt\bigl(\text{`IntAtt'}, \mathrm{H}(\sigma, \textit{trigger}, j, m_i)\bigr)$\;
\textbf{Assertion:} branch stopped at position $j$, action at $j$ produced no side effect, and no further node is committed in that branch\;
\end{attdesc}

\noindent\textbf{\underline{Sub-task Dispatch:}} This binds the planner's decomposition to the branches it spawned, so that a dropped or injected sub-agent is distinguishable from a legitimate parallel run.
It takes the sub-tasks and their per-branch trajectories, and outputs the \texttt{agent\_step} that records the dispatch.
A verifier concludes the exact sub-agents that were dispatched and their trajectories, and that the branch set is the one the parent binds through the structure root (\Cref{att:disp}).

\begin{attdesc}[htb]
\caption{Sub-task Dispatch}\label{att:disp}
\textbf{Input:} sub-tasks $\tau_1,\dots,\tau_m$ of decomposed task, with trajectories $\traj_1,\dots,\traj_m$ from APM\;
\textbf{Computation:} planner commits \texttt{agent\_step} and dispatches one sub-agent per sub-task, each rooted at \texttt{agent\_branch}\;
\textbf{Output:} $\signatt\bigl(\text{`DispAtt'}, \mathrm{H}(\tau_1,\dots,\tau_m, \traj_1,\dots,\traj_m)\bigr)$\;
\textbf{Assertion:} exactly $m$ sub-agents were dispatched, sub-agent $k$ is constrained by $\traj_k$, and complete branch set is bound to parent\;
\end{attdesc}

\noindent\textbf{\underline{Execution Graph:}} \method produces one attestation for the entire AEG, and it is the only hardware invocation during execution (the APM is attested separately, before the run starts).
It binds the recorded graph to the hardware that produced it, turning a graph anyone could recompute into evidence that this TD actually ran it.
It takes the committed nodes in commit order together with their topology, and outputs a single hardware attestation over the AEG summary $(c_N,s)$ and the public key that signed every per-node attestation in the graph.
A verifier concludes that the serialized AEG is exactly what the TD ran. Any omission, reordering, or modification of a node changes $c_N$, and any change to the topology changes $s$, and every per-node signature traces back to the certified key (\Cref{att:aeg}).
As the head and the structure root already commit to every node and the topology, this one attestation covers all the attestations above (\S\ref{sec:aeg}).

\begin{attdesc}[t]
\caption{Attested Execution Graph}\label{att:aeg}
\textbf{Input:} committed nodes with measurements $m_1,\dots,m_N$ in commit order, graph topology (each node's kind, parent, dispatched branch set), public part $\mathit{pk}$ of in-TEE signing key\;
\textbf{Computation:} fold chain $c_i = \mathrm{H}(c_{i-1}\,\|\,\textit{seq}_i\,\|\,m_i)$ to head $c_N$, compute structure root $s$ over topology, and request one hardware attestation over $(c_N,s)$ and $\mathit{pk}$\;
\textbf{Output:} AEG, $\signatt\bigl(\text{`AEGAtt'}, \mathrm{H}(c_N, s, \mathit{pk})\bigr)$, with GPU evidence binding models that ran\;
\textbf{Assertion:} serialized AEG is what this TD ran: $c_N$ binds every node's content and order, $s$ binds topology, and $\mathit{pk}$ certifies key that signed every per-node attestation\;
\end{attdesc}


\noindent\textbf{\underline{Nodes without a Separate Attestation:}} The remaining node types carry measurements but no signed statement of their own, since an existing attestation already binds them.
The \texttt{prompt} is bound by \Cref{att:apm}, a \texttt{blocked} node by the check it hangs off (\Cref{att:check}), and \texttt{tool\_output} and \texttt{final output} nodes by the chain in \Cref{att:aeg}.
Measuring a \texttt{tool\_output} records \emph{what} the executor read, which is what a verifier needs to replay a deviation; it asserts nothing about the content, since tool outputs are untrusted and are exactly where injected instructions arrive.

\end{document}